\documentclass[aps,prd,twocolumn,10pt]{revtex4-1}
\usepackage{amssymb}
\usepackage{graphicx}
\usepackage{amsmath}
\usepackage{hyperref}
\usepackage{subfigure}
\usepackage{multirow}
\usepackage{setspace}
\usepackage{verbatim}
\usepackage{float}
\usepackage{color}
\usepackage{ulem}
\usepackage[utf8]{inputenc}

\begin{document}

\title{Hydrodynamic backreaction force of cosmological bubble expansion}

\author{Shao-Jiang Wang$^{1}$}
\email{schwang@itp.ac.cn}

\author{Zi-Yan Yuwen$^{1,2}$}
\email{yuwenziyan@itp.ac.cn}

\affiliation{$^1$CAS Key Laboratory of Theoretical Physics, Institute of Theoretical Physics, Chinese Academy of Sciences, Beijing 100190, China}
\affiliation{$^2$School of Physical Sciences, University of Chinese Academy of Sciences (UCAS), Beijing 100049, China}
%\date{\today}

\begin{abstract}
As a promising probe for the new physics beyond the standard model of particle physics in the early Universe, the predictions for the stochastic gravitational wave background from a cosmological first-order phase transition heavily rely on the bubble wall velocity determined by the bubble expansion dynamics. The bubble expansion dynamics is governed by the competition between the driving force from the effective potential difference and the backreaction force from a sum of the thermal force and friction force induced by the temperature jumping and out-of-equilibrium effects across the bubble wall, respectively. In this paper, we propose a hydrodynamic evaluation on this total backreaction force for a non-runaway steady-state bubble expansion, which, after evaluated at the wall interface, exactly reproduces the pressure difference $\Delta_\mathrm{wall}[(\bar{\gamma}^2-1)w]$ obtained previously from the junction condition of the total energy-momentum tensor at the wall interface, where $w$ is the enthalpy and $\bar{\gamma}\equiv(1-\bar{v}^2)^{-1/2}$ is the Lorentz factor of the wall-frame fluid velocity $\bar{v}$.
\end{abstract}
\maketitle

\section{Introduction}

One intriguing possibility for going beyond the standard model (BSM) of particle physics in the early Universe is the presence of the cosmological first-order phase transitions (FOPTs)~\cite{Mazumdar:2018dfl,Hindmarsh:2020hop,Caldwell:2022qsj}. Besides the stochastic gravitational wave backgrounds (SGWBs) ~\cite{Caprini:2015zlo,Caprini:2019egz,Cai:2017cbj,Bian:2021ini}, the production of primordial black holes is also a universal consequence of the cosmological FOPTs~\cite{Liu:2021svg,Hashino:2021qoq} (see also~\cite{Baker:2021nyl,Baker:2021sno,Kawana:2021tde,Huang:2022him,Marfatia:2021hcp} for other specific mechanisms). The associated non-equilibrium nature during FOPT can also be used to explain the baryon asymmetry~\cite{Morrissey:2012db} and primordial magnetic fields~\cite{Subramanian:2015lua}. Therefore, the cosmological FOPT can serve as a promising probe for the BSM new physics in the early Universe.

The cosmological FOPT starts with stochastic nucleations of true vacuum bubbles in the false vacuum background filled with the thermal plasma. The true vacuum bubble is a scalar field profile interpolating the outside false and inside true vacua separated by a barrier in the effective potential. The effective potential mimics the free energy density as the opposite of the pressure acting on the bubble wall. Since the effective potential is higher in the false vacuum than in the true vacuum, the pressure difference induced by the effective potential pushes the wall outwards to encounter with the plasma, which in turn imposes a backreaction on the bubble wall.

If this backreaction force has no explicit dependence on the Lorentz factor $\gamma_w=1/\sqrt{1-v_w^2}$  of the bubble wall velocity $v_w$, then the bubble wall simply accelerates to approach the speed of light.  This  runaway expansion would contribute to the runaway collisions, of which the SGWB is dominated by bubble wall collisions. If this backreaction force possesses some $\gamma_w^n$-dependence for a positive $n$, then the bubble wall would eventually approach a steady state with a terminal wall velocity that balances the driving force and backreaction force. This non-runaway expansion can still contribute to runway collisions if bubbles can collide with each other even before they could ever close to the terminal wall velocity,  where the SGWB is still dominated by bubble wall collisions.  Otherwise, it is dominated by sound waves.

One contribution to the total backreaction force is the friction force from the out-of-equilibrium effects. The understanding of the $\gamma_w$-dependence for the friction force has been evolved and shaped with twists and turns over the past decade~\cite{Bodeker:2009qy,Bodeker:2017cim,Hoeche:2020rsg,Gouttenoire:2021kjv} until recently it becomes an active yard of hot debates \cite{BarrosoMancha:2020fay,Vanvlasselaer:2020niz,Balaji:2020yrx,Ai:2021kak,Dorsch:2021nje,DeCurtis:2022hlx}\cite{Bea:2021zsu,Bigazzi:2021ucw}. 
Nevetherless, regardless of the detailed $\gamma_w$-dependence in the friction force, the efficiency factor of inserting the released vacuum energy into the bubble wall kinetic energy can be calculated in general from an effective picture of the bubble expansion~\cite{Cai:2020djd}. The other contribution to the backreaction force is recognized recently in Ref.~\cite{Ai:2021kak} as the thermal force from the temperature jumping across the bubble wall, which is usually overlooked in the literature except for some earlier studies~\cite{Espinosa:2010hh,Konstandin:2010dm} that attributes it to a modification on the driving force. 

In this paper, we propose a hydrodynamic formulation on the total backreaction force consisting of the thermal force and friction force for a non-runaway steady-state bubble expansion without assuming a bag equation of state (EoS) and without requiring an equilibrium distribution function. For an illustrative example with a bag EoS, we can specifically evaluate the total backreaction force that exactly balances the driving force. In particular, when evaluated at the wall interface, our hydrodynamic evaluation on the backreaction force exactly reproduces the pressure difference $\Delta_\mathrm{wall}[(\bar{\gamma}^2-1)w]$ obtained previously from the junction condition of the total energy-momentum tensor at the wall interface, where $w$ is the enthalpy and $\bar{\gamma}\equiv(1-\bar{v}^2)^{-1/2}$ is the Lorentz factor of the wall-frame fluid velocity $\bar{v}$. This provides a new hydrodynamic perspective into the bubble expansion dynamics.

%In particular, for an illustrative case with a bag equation of state, our hydrodynamic evaluation on the pressure difference $\Delta[(\bar{\gamma}^2-1)w]$ over the whole sound-shell part with a non-vanishing fluid velocity profile exactly reproduces the leading $(\gamma_w^2-1)$-behavior from a previous field-theoretic evaluation on the pressure difference $\Delta[(\bar{\gamma}^2-1)w]$ away from the bubble wall in both local thermal equilibrium and ballistic approximations, where $\gamma_w$ is the Lorentz factor of the bubble wall velocity.

%It seems that the thermal force could be obtained from the total backreaction force when assuming a thermal equilibrium in the vicinity of the bubble wall, so that the junction condition at the wall interface would give rise to a pressure difference of form $\Delta[(\bar{\gamma}^2-1)w]$~\cite{BarrosoMancha:2020fay,Balaji:2020yrx}, where $w$ is the enthalpy and $\bar{\gamma}\equiv1/\sqrt{1-\bar{v}^2}$ is the Lorentz factor of the fluid velocity $\bar{v}$ in the wall frame. However, $\Delta[(\bar{\gamma}^2-1)w]$ is only derived at the wall interface with a local thermal equilibrium but neglects all those effects from the sound shell defined by the non-vanishing part of the fluid velocity profile. In this paper, we propose a complete hydrodynamic evaluation on the backreaction forces, which not only recovers $\Delta[(\bar{\gamma}^2-1)w]$ at the wall interface but also exactly matches the direct calculation on the backreaction force for the case with a bag equation of state (EoS).

\section{Scalar-plasma system}

We start with the scalar-plasma system with energy-momentum tensors for the scalar field and thermal plasma (neglecting the shear and the bulk viscosity) of usual forms
\begin{align}
T_\phi^{\mu\nu}&=\nabla^\mu\phi\nabla^\nu\phi+g^{\mu\nu}\left[-\frac12(\nabla\phi)^2-V_0(\phi)\right],\\
T_f^{\mu\nu}&=\sum\limits_{i=\mathrm{B,F}}g_i\int\frac{\mathrm{d}^3\mathbf{k}}{(2\pi)^3}\frac{k^\mu k^\nu}{k^0}\bigg|_{k^0=E_i(\mathbf{k})}f_i(\mathbf{x},\mathbf{k}),\label{eq:Tf}
\end{align}
respectively, where $E_i(\mathbf{k})\equiv\sqrt{\mathbf{k}^2+m_i^2}$ is the energy of the particle of species $i$ with momentum $\mathbf{k}$ and effective mass $m_i$, and $f_i(\mathbf{x},\mathbf{k})$ is the distribution function counting the average number of particles of species $i$ with the momentum $\mathbf{k}$ and energy $E_i(\mathbf{k})$ in a volume element $(\mathbf{x},\mathbf{x}+\mathrm{d}\mathbf{x})\times(\mathbf{k},\mathbf{k}+\mathrm{d}\mathbf{k})$ of the phase space at the time $t=x^0$. Here the total effective potential $V_\mathrm{eff}=V_0(\phi)+V_T^\mathrm{1-loop}+\Delta V_T\equiv V_0+V_T$ contains a zero-temperature part $V_0(\phi)$, a finite-temperature part $V_T^\mathrm{1-loop}$ at one loop, and a finite-temperature part $\Delta V_T$ from higher loops.   

The conservation of the total energy-momentum tensor $\nabla_\mu(T_\phi^{\mu\nu}+T_f^{\mu\nu})=0$ requires
\begin{align}
\nabla_\mu T_\phi^{\mu\nu}&\equiv[\nabla_\mu\nabla^\mu\phi-V'_0(\phi)]\nabla^\nu\phi=+f^\nu,\\
\nabla_\mu T_f^{\mu\nu}&\equiv\sum\limits_{i=\mathrm{B,F}}g_i\int\frac{\mathrm{d}^3\mathbf{k}}{(2\pi)^3}\frac{k^\mu k^\nu}{E_i(\mathbf{k})}\nabla_\mu f_i=-f^\nu,
\end{align}
where the transfer flow $f^\nu$ can be determined as
\begin{align}
f^\nu=\nabla^\nu\phi\sum\limits_{i=\mathrm{B,F}}g_i\frac{\mathrm{d}m_i^2}{\mathrm{d}\phi}\int\frac{\mathrm{d}^3\mathbf{k}}{(2\pi)^3}\frac{f_i}{2E_i}
\end{align}
from integrating the Boltzmann equation~\cite{Moore:1995ua,Moore:1995si} (see also~\cite{Hindmarsh:2020hop}). Note that there comes a simple relation
\begin{align}
\frac{\partial V_T^\mathrm{1-loop}}{\partial \phi}=\sum\limits_{i=\mathrm{B,F}}g_i\frac{\mathrm{d}m_i^2}{\mathrm{d}\phi}\int\frac{\mathrm{d}^3\mathbf{k}}{(2\pi)^3}\frac{f_i^\mathrm{eq}}{2E_i}
\end{align}
between the thermal equilibrium distribution functions
\begin{align}
f_i^\mathrm{eq}(\mathbf{x},\mathbf{k})=\frac{1}{e^{(E_i(\mathbf{k})-\mu_i)/T}\mp1}
\end{align}
from the Bose-Einstein/Fermi-Dirac distributions with negligible chemical potentials and the one-loop finite-temperature potential
\begin{align}
V_T^\mathrm{1-loop}=\sum_{i=\mathrm{B,F}}\pm g_i T\int\frac{\mathrm{d}^3\mathbf{k}}{(2\pi)^3}\log\left(1\mp e^{-\frac{E_i(\mathbf{k})}{T}}\right).
\end{align}

Then, we can split the total distribution function into $f_i=f_i^\mathrm{eq}+\Delta f_i^\mathrm{eq}+\delta f_i$ so that the transfer flow can be parametrized as
\begin{align}
f^\nu=\nabla^\nu\phi\left(\frac{\partial V_T^\mathrm{1-loop}}{\partial \phi}+\frac{\partial \Delta V_T}{\partial \phi}-\frac{\partial p_{\delta f}}{\partial\phi}\right)
\end{align}
corresponding to the leading-order equilibrium part, higher-order equilibrium part, and non-equilibrium part, respectively. As we will see later, this parametrization splitting is crucial to obtain the usual conservation equation of entropy flow that can be identically vanished when the out-of-equilibrium term $\partial p_{\delta f}/\partial\phi$ is absent. Therefore, the scalar and plasma parts of Boltzmann equation of motions (EoMs) becomes
\begin{align}
\nabla_\mu\nabla^\mu\phi-\frac{\partial V_\mathrm{eff}}{\partial\phi}&=-\frac{\partial p_{\delta f}}{\partial\phi},\label{eq:EOMscalar}\\
\nabla_\mu T_f^{\mu\nu}+\nabla^\nu\phi\frac{\partial V_T}{\partial\phi}&=\nabla^\nu\phi\frac{\partial p_{\delta f}}{\partial\phi},\label{eq:EOMplasma}
\end{align}
as expected also from the Kadanoff-Baym equations~\cite{Konstandin:2014zta}.

\section{Bubble-fluid system}

For a fast FOPT completed within one Hubble time, the Hubble expansion can be neglected, and we work in the Minkowski spacetime below with the origin fixed at the bubble center. For bubbles nucleated with a negligible initial size, the spherical bubble expansion is self-similar in such a way that the scalar profile $\phi(t,r)$ evolves as a function of the self-similar coordinate $\xi\equiv r/t$ alone. For a non-runaway expansion quickly reaching a steady state with a terminal wall velocity $v_w$, a thin-wall scalar profile $\phi(t,r)\equiv\phi(\xi)=\phi_+\Theta(\xi-v_w)+\phi_-\Theta(v_w-\xi)$ is assumed so that $\phi'(\xi)=(\phi_+-\phi_-)\delta(\xi-v_w)$. In a local frame comoving with the bubble wall in the $z$ direction, for example, the bubble wall can be treated as a planer wall. Then the Lorentz transformations $\bar{t}\equiv\gamma_w(t-v_wz)$ and $\bar{z}\equiv\gamma_w(z-v_wt)$ lead to a vanishing time derivative for the scalar profile in the local wall frame, $\partial_{\bar{t}}\phi=(\gamma_w/t)(v_w-\xi)\phi'(\xi)=(\gamma_w/t)(v_w-\xi)(\phi_+-\phi_-)\delta(\xi-v_w)=0$. Therefore, in the local wall frame, $T_\phi^{\mu\nu}$ adimits non-vanishing components $T_\phi^{\bar{t}\bar{t}}= V_0$ and $T_\phi^{\bar{z}\bar{z}}=-V_0$. Reversing back to the background plasma frame via the Lorentz transformations  $T_\phi^{\bar{t}\bar{t}}=\gamma_w^2(T_\phi^{tt}+T_\phi^{zz})-T_\phi^{zz}$ and $T_\phi^{\bar{z}\bar{z}}=(\gamma_w^2-1)(T_\phi^{tt}+T_\phi^{zz})+T_\phi^{zz}$ simply yields $T_\phi^{tt}=T_\phi^{\bar{t}\bar{t}}=V_0\equiv e_\phi$ and $T_\phi^{zz}=T_\phi^{\bar{z}\bar{z}}=-V_0\equiv p_\phi$, which leads to a perfect fluid form $T_\phi^{\mu\nu}=(e_\phi+p_\phi)u^\nu u^\nu+p_\phi\eta^{\mu\nu}$ for the scalar field with the enthalpy $w_\phi=e_\phi+p_\phi=0$.

The energy-momentum tensor~\eqref{eq:Tf} for the thermal plasma can be also of a perfect fluid form,
\begin{align}\label{eq:plasmafluid}
T_f^{\mu\nu}
&=\sum\limits_{i=\mathrm{B,F}}g_i\int\frac{\mathrm{d}^3\mathbf{k}}{(2\pi)^3}\left[\left(E_i+\frac{\mathbf{k}^2}{3E_i}\right)u^\mu u^\nu+\eta^{\mu\nu}\frac{\mathbf{k}^2}{3E_i}\right]f_i\nonumber\\
&\equiv(e_f+p_f)u^\mu u^\nu+p_f\eta^{\mu\nu},
\end{align}
provided that the plasma energy density $e_f$ and pressure $p_f$ are recognized locally in the plasma rest frame $u^\mu=(1,0,0,0)$ comoving with the fluid elements as
\begin{align}
e_f&=\sum\limits_{i=\mathrm{B,F}}g_i\int\frac{\mathrm{d}^3\mathbf{k}}{(2\pi)^3}E_i(\mathbf{k})f_i,\\
p_f&=\sum\limits_{i=\mathrm{B,F}}g_i\int\frac{\mathrm{d}^3\mathbf{k}}{(2\pi)^3}\frac{\mathbf{k}^2}{3E_i(\mathbf{k})}f_i.
\end{align}
It is worth noting that although the perfect fluid form for~\eqref{eq:Tf} does not necessarily require the distribution function $f_i$ to be exactly the equilibrium one $f_i^\mathrm{eq}$, the local equilibrium is implicitly assumed that the space-time dependence in $f_i$ is reduced into a single self-similar coordinate $\xi\equiv r/t$ so that both the energy density and pressure as well as the associated temperature profiles are spherically homogeneous at a given $\xi$.
The fluid velocity in the background plasma frame reads $u^\mu=\gamma(1,v,0,0)$ in a spherical coordinate system. For the fluid velocity $\bar{v}$ in a local wall frame moving along $z$ direction, it holds $u^\mu=\bar{\gamma}(1,0,0,-\bar{v})$ with $\bar{\gamma}=(1-\bar{v})^{-1/2}$ and  $-\bar{v}=(v-v_w)/(1-v v_w)$, where the minus sign in front of $\bar{v}$ is introduced to ensure a positive $\bar{v}$ for later convenience. Therefore, in the local wall frame, $T_f^{\mu\nu}$ admits non-vanishing components $T_f^{\bar{t}\bar{t}}=w\bar{\gamma}^2-p_f$, $T_f^{\bar{t}\bar{z}}=T_f^{\bar{z}\bar{t}}=-w\bar{\gamma}^2\bar{v}$ and $T_f^{\bar{z}\bar{z}}=w\bar{\gamma}^2\bar{v}^2+p_f$, where we have used the fact that $w=w_\phi+w_f=w_f=e_f+p_f$.

In summary, the total energy-momentum tensor therefore also admits a perfect fluid form $T^{\mu\nu}=(e+p)u^\mu u^\nu+p\eta^{\mu\nu}$ with $e=e_f+e_\phi=e_f+V_0$, $p=p_f+p_\phi=p_f-V_0$. Note that $-p=\mathcal{F}=V_\mathrm{eff}=V_0+V_T$, then $p_f=p+V_0=-V_\mathrm{eff}+V_0=-V_T$.
The conservation of the total energy-momentum tensor $\nabla_\mu(T_\phi^{\mu\nu}+T_f^{\mu\nu})=0$ at the wall interface in a local wall frame then induces junction conditions
\begin{align}
\partial_{\bar{z}}T^{\bar{z}\bar{t}}=0&\Rightarrow w_-\bar{\gamma}_-^2\bar{v}_-=w_+\bar{\gamma}_+^2\bar{v}_+,\label{eq:1stjunction}\\
\partial_{\bar{z}}T^{\bar{z}\bar{z}}=0&\Rightarrow w_-\bar{\gamma}_-^2\bar{v}_-^2+p_-=w_+\bar{\gamma}_+^2\bar{v}_+^2+p_+,\label{eq:2ndjunction}
\end{align}
with subscripts ``$\pm$'' for false/true vacua just right outside and inside of the bubble wall interface, respectively.

\section{Backreaction force}

The physical roles played by the dubbed driving force and backreaction force are more physically motivated from  integrating the Boltzmann EoM ~\eqref{eq:EOMscalar} of the scalar field across the bubble wall,
\begin{align}
\int\mathrm{d}\xi\frac{\mathrm{d}\phi}{\mathrm{d}\xi}\left(\nabla^2\phi-\frac{\partial V_\mathrm{eff}}{\partial\phi}+\frac{\partial p_{\delta f}}{\partial\phi}\right)=0.
\end{align}
The first term vanishes as a total derivative of $\phi'(\xi)^2$, the second term could be split into two contributions by $(\partial V_\mathrm{eff}/\partial\phi)(\mathrm{d}\phi/\mathrm{d}\xi)=\mathrm{d}V_\mathrm{eff}/\mathrm{d}\xi-(\partial V_\mathrm{eff}/\partial T)(\mathrm{d}T/\mathrm{d}\xi)$ if one fully appreciates the temperature jumping across the bubble wall~\cite{Espinosa:2010hh,Konstandin:2010dm}. The outcome,
\begin{align}\label{eq:balance}
\frac{F_\mathrm{drive}}{A}\equiv\Delta V_\mathrm{eff}&=\int\mathrm{d}\xi\frac{\mathrm{d}T}{\mathrm{d}\xi}\frac{\partial V_\mathrm{eff}}{\partial T}+\int\mathrm{d}\xi\frac{\mathrm{d}\phi}{\mathrm{d}\xi}\frac{\partial p_{\delta f}}{\partial\phi}\nonumber\\
&\equiv\frac{F_\mathrm{themo}}{A}+\frac{F_\mathrm{fric}}{A}\equiv\frac{F_\mathrm{back}}{A},
\end{align}
can be interpreted as a balance between the driving force and the backreaction force consisting of the thermal force caused by the temperature jumping from $\mathrm{d}T/\mathrm{d}\xi$ and the friction force caused by the out-of-equilibrium effect from $\delta f_i$. Here $A$ is the area of a region on the bubble wall that can be approximately treated as a planar wall.

\subsection{Backreaction force at the wall interface}

It is easy to see from the second junction condition~\eqref{eq:2ndjunction} that the pressure difference $\Delta_\mathrm{wall}(p)\equiv p_+-p_-=w_-\bar{\gamma}_-^2\bar{v}_-^2-w_+\bar{\gamma}_+^2\bar{v}_+^2\equiv-\Delta_\mathrm{wall}(\bar{\gamma}^2\bar{v}^2w)$ can be used to indirectly evaluate the backreaction force acting on the wall interface,
\begin{align}\label{eq:balanceatwall}
\frac{F_\mathrm{back}^\mathrm{wall}}{A}&=\Delta_\mathrm{wall}(V_\mathrm{eff})=-\Delta_\mathrm{wall}(p)=\Delta_\mathrm{wall}(\bar{\gamma}^2\bar{v}^2w),
\end{align}
through the driving force $F_\mathrm{drive}/A\equiv\Delta V_\mathrm{eff}=F_\mathrm{back}/A$ that balances the backreaction force for a non-runaway steady-state bubble expansion. It would be appealing to directly evaluate the full backreaction force and its contribution at the wall interface alone as shown shortly below. It is worth noting that, the difference of $\bar{\gamma}^2\bar{v}^2w\equiv(\bar{\gamma}^2-1)w$ taken just right outside and inside of the wall, 
\begin{align}
\Delta_\mathrm{wall}(\bar{\gamma}^2\bar{v}^2w)
&\equiv\lim_{\delta\to0}\bigg[(\bar{\gamma}(\bar{v}(\xi))^2-1)w(\xi)\bigg]\bigg|_{\xi=v_w-\delta}^{\xi=v_w+\delta}\nonumber\\
&\neq(\gamma_w^2-1)\Delta_\mathrm{wall}w,
\end{align}
does not reproduce the $(\gamma_w^2-1)$ factor as we will check explicitly with hydrodynamics, in which the detonation case, for example, admits $\bar{v}_+=v_w$ but $\bar{v}_-\neq v_w$.  In fact, due to $\bar{v}(\xi=0,1)=v_w$, it is the expression
\begin{align}
\Delta(\bar{\gamma}^2\bar{v}^2w)&\equiv\bigg[(\bar{\gamma}^2-1)w\bigg]\bigg|_{\xi=0}^{\xi=1}=(\gamma_w^2-1)\Delta w
\end{align}
that holds true for the difference of $(\bar{\gamma}^2-1)w$ taken sufficiently distant away from the bubble wall where the thermal equilibrium could be approximately established. The resulted $(\gamma_w^2-1)$-behavior is exactly what was found in Ref.~\cite{BarrosoMancha:2020fay} (see also~\cite{Balaji:2020yrx}) as the leading contribution to a field-theoretic evaluation on the pressure difference away from the bubble wall in both cases with the local thermal equilibrium and ballistic approximations. However, the difference $\Delta(\bar{\gamma}^2\bar{v}^2w)$ cannot be used to evaluate the full backreaction force (not just its contribution at the wall interface) simply because that the last equality in Eq.~\eqref{eq:balanceatwall} is only valid at the wall interface.

\subsection{Thermal force}

The thermal force is arisen due to the temperature jumping across the bubble wall, and can be always expressed formally as
\begin{align}\label{eq:themo}
\frac{F_\mathrm{themo}}{A}
&\equiv\int\mathrm{d}\xi\frac{\mathrm{d}T}{\mathrm{d}\xi}\frac{\partial V_\mathrm{eff}}{\partial T}\nonumber\\
&=\int\mathrm{d}\xi\frac{\mathrm{d}T}{\mathrm{d}\xi}\left(\frac{\partial\epsilon}{\partial T}-\frac13\frac{\partial a}{\partial T}T^4-\frac43 a(T) T^4\right)\nonumber\\
&=\int\mathrm{d}\xi\frac{\mathrm{d}T}{\mathrm{d}\xi}\left(-\frac43a(T)T^3\right)\equiv-\int s(T)\mathrm{d}T
\end{align}
even beyond the bag EoS as long as we can formally parametrize  the  pressure and energy density as
\begin{align}
p(T)=\frac13a(T)T^4-\epsilon(T),\quad
e(T)=a(T)T^4+\epsilon(T)
\end{align}
in terms of some bag-like quantities 
\begin{align}
a(T)\equiv\frac{3}{4T^3}\frac{\partial p}{\partial T}=\frac{3w}{4T^4}, \quad
\epsilon(T)\equiv\frac{e(T)-3p(T)}{4}
\end{align}
that mimic the bag EoS ones so that a relation $\partial\epsilon/\partial T=(T^4/3)(\partial a/\partial T)$ can be always satisfied for a cancellation in arriving at~\eqref{eq:themo}. Here $s(T)\equiv w(T)/T=[e(T)+p(T)]/T=(4/3)a(T)T^3$ is the entropy density.

\subsection{Friction force}

The friction force comes from the out-of-equilibrium term $\partial p_{\delta f}/\partial\phi$, which can be formulated with the help of hydrodynamics. We first insert the perfect fluid ansatz~\eqref{eq:plasmafluid} into the Boltzmann EoM~\eqref{eq:EOMplasma} of the thermal plasma, and then multiply both sides with $u_\nu$,
\begin{align}
u_\nu\nabla_\mu(wu^\mu u^\nu+p_f\eta^{\mu\nu})+u_\nu\nabla^\nu\phi\frac{\partial V_T}{\partial\phi}=u_\nu\nabla^\nu\phi\frac{\partial p_{\delta f}}{\partial\phi}.
\end{align}
After using the relations $u_\nu u^\nu=-1$, $u_\nu\nabla_\mu u^\nu=0$, and $\nabla_\mu p_f=-\nabla_\mu V_T=-\nabla_\mu T(\partial V_T/\partial T)-\nabla_\mu\phi(\partial V_T/\partial\phi)=-\nabla_\mu T(\partial V_\mathrm{eff}/\partial T)-\nabla_\mu\phi(\partial V_T/\partial\phi)$, it becomes a conservation equation for the enthalpy flow,
\begin{align}\label{eq:enthalpyflow}
-\nabla_\mu(wu^\mu)=u^\mu\nabla_\mu T\frac{\partial V_\mathrm{eff}}{\partial T}+u^\mu\nabla_\mu\phi\frac{\partial p_{\delta f}}{\partial\phi}.
\end{align}
At this point, if we expand  $\nabla_\mu(wu^\mu)=T\nabla_\mu(su^\mu)+(su^\mu)\nabla_\mu T$ and then cancel out the last term with the first term $u^\mu\nabla_\mu T(\partial V_\mathrm{eff}/\partial T)=-u^\mu\nabla_\mu T(\partial p/\partial T)=-u^\mu(\nabla_\mu T)s$ on the right-hand-side (RHS), one recovers the usual conservation equation of the entropy flow,  
\begin{align}\label{eq:entropyflow}
T\nabla_\mu(su^\mu)=-u^\mu\nabla_\mu\phi\frac{\partial p_{\delta f}}{\partial\phi},
\end{align}
which is violated as expected by the out-of-equilibrium effect contributing exactly at the wall interface due to the presence of the prefactor $u^\mu\nabla_\mu\phi=(\gamma,\gamma v,0,0)\cdot(-\xi/t,1/t,0,0)\partial_\xi\phi=(\gamma/t)(v-\xi)\phi'(\xi)$ that is vanished everywhere except for the wall interface. In fact, we can explicitly expand $\nabla_\mu(su^\mu)$ in the plasma frame as
\begin{align}
\nabla_\mu(su^\mu)&=\partial_t(su^t)+\frac{2}{r}(su^r)+\partial_r(su^r)\nonumber\\
&=-\frac{\xi}{t}\partial_\xi(s\gamma)+\frac{2}{r}(s\gamma v)+\frac{1}{t}\partial_\xi(s\gamma v)\nonumber\\
&=\frac{\gamma}{t}(v-\xi)\partial_\xi s+\frac{2}{\xi}\frac{\gamma}{t}sv+\frac{\gamma^3}{t}s(1-\xi v)\partial_\xi v,
\end{align}
which, after replacing for the $\mathrm{d}w/\mathrm{d}\xi$ and $2v/\xi$ terms with the fluid EoMs~\eqref{eq:dv} and~\eqref{eq:dw},  renders the entropy flow to be conserved everywhere except for the wall interface,
\begin{align}
\frac{t}{\gamma}\nabla_\mu(su^\mu)&=(v-\xi)\partial_\xi\left(\frac{w}{T}\right)+\frac{w}{T}\left[\frac{2v}{\xi}+\gamma^2(1-\xi v)\partial_\xi v\right]\nonumber\\
&=\frac{w}{T}\frac{\gamma^2}{c_s^2}(\mu\partial_\xi v)[(v-\xi)+(1-\xi v)\mu]=0.
\end{align}
This is different from the Ref.~\cite{Ai:2021kak} where the entropy flow is taken to be conserved by assuming a local equilibrium at the wall interface with a vanishing out-of-equilibrium term $\partial p_{\delta f}/\partial\phi=0$ and hence a vanishing friction force. In this case, the temperature profile cannot stay constant across the bubble wall, otherwise the thermal force~\eqref{eq:themo} is also vanished, leading to a vanishing total backreaction force that cannot balance the driving force for a steady wall expansion. This is why Ref.~\cite{Ai:2021kak} must find an inhomogeneous temperature profile and subsequent findings therein when assuming a local equilibrium at the wall interface. However, even with the presence of the out-of-equilibrium term, the entropy-flow equation has already suggested a hydrodynamic expression for the friction force as
\begin{align}\label{eq:frichydro}
\frac{F_\mathrm{fric}^\mathrm{hydro}}{A}
&=\int\mathrm{d}\xi\frac{\mathrm{d}\phi}{\mathrm{d}\xi}\frac{\partial p_{\delta f}}{\partial\phi}=\int\mathrm{d}\xi\frac{Tt\nabla_\mu(su^\mu)}{\gamma(\xi-v)}\nonumber\\
&=\int_0^1\mathrm{d}\xi\left(-T\frac{\mathrm{d}s}{\mathrm{d}\xi}+\frac{2wv}{\xi(\xi-v)}+\frac{w\gamma^2}{\mu}\frac{\mathrm{d}v}{\mathrm{d}\xi}\right).
\end{align}

\subsection{Full backreaction force}

It is worth noting that the RHS of the enthalpy-flow Eq.~\eqref{eq:enthalpyflow} already reproduces the integrand of the RHS of the balance Eq.~\eqref{eq:balance}, therefore, the enthalpy-flow equation can also lead to a hydrodynamic expression for the total backreaction force as
\begin{align}\label{eq:backhydro}
\frac{F_\mathrm{back}^\mathrm{hydro}}{A}=\int\mathrm{d}\xi\frac{t\nabla_\mu(wu^\mu)}{\gamma(\xi-v)}.
\end{align}
Note that in deriving our hydrodynamic expression~\eqref{eq:backhydro} for the total backreaction force, we assume neither a constant temperature profile across the bubble wall nor a vanishing out-of-equilibrium term at the wall interface. To actually evaluate the total backraction force,  we can explicitly expand $\nabla_\mu(wu^\mu)$ in the plasma frame as
\begin{align}
\nabla_\mu(wu^\mu)=\frac{\gamma}{t}(v-\xi)\partial_\xi w+\frac{2}{\xi}\frac{\gamma}{t}wv+\frac{\gamma^3}{t}w(1-\xi v)\partial_\xi v,
\end{align}
so that the integral becomes
\begin{align}\label{eq:FbackHydro}
\frac{F_\mathrm{back}^\mathrm{hydro}}{A}=\int_0^1\mathrm{d}\xi\left(-\frac{\mathrm{d}w}{\mathrm{d}\xi}+\frac{2wv}{\xi(\xi-v)}+\frac{w\gamma^2}{\mu}\frac{\mathrm{d}v}{\mathrm{d}\xi}\right)
\end{align}
with an abbreviation $\mu(\xi,v)\equiv(\xi-v)/(1-\xi v)$. This explicit hydrodynamic expression can be further split into a sound-shell part with a non-vanishing fluid velocity and discontinuous parts consisting of the wall interface and shockwave front, if any. For the sound shell part,  the terms involving with $v/\xi$ and $\mathrm{d}v/\mathrm{d}\xi$ can be replaced by the fluid EoMs~\eqref{eq:dv} and~\eqref{eq:dw} so that the sound shell contribution can be explicitly worked out as
\begin{align}\label{eq:FbackShell}
\frac{F_\mathrm{back}^\mathrm{hydro}}{A}\bigg|_\mathrm{shell}=-\int_\mathrm{shell}\mathrm{d}\xi\frac{\mathrm{d}w}{\mathrm{d}\xi}\frac{c_s^2}{1+c_s^2},
\end{align}
where the sound speed is formally defined as $c_s^2=\partial p/\partial e$, and the integration over the sound shell has excluded the parts from the bubble wall and shock front (if any). For the discontinuous parts, only the first and last terms in the integrand contribute to our hydrodynamic evaluation on the total backreaction force but in a rather non-trivial manner as we will elaborate shortly below.

\subsection{Bubble-wall and shock-front contributions}

The discontinuous contributions to our explicit hydrodynamic expression~\eqref{eq:FbackHydro}  on the total backreaction force involve with derivative terms like $\mathrm{d}w/\mathrm{d}\xi$ and $\mathrm{d}v/\mathrm{d}\xi$, which are actually Dirac delta functions since both enthalpy and fluid velocity profiles would experience some sudden changes at the wall interface or shockwave front, if any. In particular, the term involving with $\mathrm{d}v/\mathrm{d}\xi$ is rather non-trivial since the factor $w\gamma^2/\mu$ in the front of $\mathrm{d}v/\mathrm{d}\xi$  is also discontinuous at the wall interface or shockwave front, if any. From a mathematical point of view, the key point is to evaluate the integral of a form,
\begin{align}
I=\int_a^b \mathrm{d}x \, f(x) \delta(x-c),\quad a<c<b.
\end{align}
If $f(x)$ is a continuous function, this integral is simply $I=f(c)$. However, if $f(x)$ also admits a discontinuity at $x=c$, then we should also find a discontinuous function $g(x)$ at $x=c$ so that $g'(x)=\delta(x-c)$, hence 
\begin{align}
I=\int_{g(a)}^{g(b)}\mathrm{d}g(x)\,f(g(x)).
\end{align}
If there is a junction condition at $x=c$ that could relate $f$ and $g$ by a continuous function $f=h(g)$, then the integral can be computed by
\begin{align}
I=\lim_{\delta\to0}\int_{g(c-\delta)}^{g(c+\delta)}\mathrm{d}g\, h(g).
\end{align}
Therefore, the bubble-wall contribution to the full backreaction force should be computed by
\begin{align}\label{eq:FbackWall}
\frac{F_\mathrm{back}^\mathrm{hydro}}{A}\bigg|_\mathrm{wall}=-\Delta_\mathrm{wall}w+\int_{v_-}^{v_+}\mathrm{d}v\frac{w(v)\gamma(v)^2}{\mu(v_w,v)},
\end{align}
with the enthalpy difference $\Delta_\mathrm{wall}w\equiv(w_+-w_-)$ taken just in the front and back of the bubble wall. Similarly, the shock-front contribution to the full backreaction force should be computed by
\begin{align}\label{eq:FbackShock}
\frac{F_\mathrm{back}^\mathrm{hydro}}{A}\bigg|_\mathrm{shock}=-\Delta_\mathrm{shock}w+\int_{v(v_{sh})}^{0}\mathrm{d}v\frac{w(v)\gamma(v)^2}{\mu(v_{sh},v)},
\end{align}
with the enthalpy difference $\Delta_\mathrm{shock}w\equiv(w_R-w_L)$ taken just in the front and back of the shock front.

\section{Hydrodynamic evaluations}

Our explicit hydrodynamic expression~\eqref{eq:FbackHydro} for the total backreaction force is ready to be evaluated once the profiles of the enthalpy and fluid velocity  can be solved from the fluid EoMs~\eqref{eq:dv} and~\eqref{eq:dw} with an appropriate EoS-dependent form of the first junction condition~\eqref{eq:1stjunction} at the wall interface and shockwave front, if any.

As the simplest illustration to verify our explicit hydrodynamic expression~\eqref{eq:FbackHydro} for the total backreaction force, we will employ the bag EoS~\eqref{eq:bagEoS} in the first junction condition to solve for the enthalpy and fluid velocity profiles from the hydrodynamic fluid EoMs as presented in details in the appendix ~\ref{app:hydrodynamics}. With a bag EoS, the driving force simply reads
\begin{align}\label{eq:BagEoSFdrive}
\frac{F_\mathrm{drive}}{A}\equiv\Delta V_\mathrm{eff}\equiv-\Delta p&=\Delta\left(-\frac13aT^4+V_0\right)\nonumber\\
&=-\frac14\Delta w+\frac34\alpha_Nw_N,
\end{align}
where $\alpha_N=4\Delta\epsilon/3w_N$ is the strength factor  with the subscript ``$N$'' for the asymptotic value at null infinity $\xi=1$, and the enthalpy difference $\Delta w=w_N-w_O$ is taken between the null infinity and bubble center. With the sound speed $c_s^2=1/3$ for the bag EoS, the sound-shell contribution~\eqref{eq:FbackShell} to the total backreaction forces,
\begin{align}\label{eq:BagEoSFshell}
\frac{F_\mathrm{back}^\mathrm{hydro}}{A}\bigg|_\mathrm{shell}=-\frac14\Delta_\mathrm{shell} w,
\end{align}
is simply the opposite of one quarter of the enthalpy difference between boundaries of the sound shell. The discontinuous parts of the total backreaction force will be evaluated separately below for three different expansion modes with details attached in the appendix.~\ref{app:hydrodynamics}.

\subsection{Detonation}

For the detonation mode, there is only one discontinuity at the wall interface. Since both $w$ and $v$ have experienced a jump across the bubble wall, we require extra input for $w$ as a function of $v$ at the wall interface if we take $v$ as the integration variable, which can be achieved by the junction condition
\begin{align}
w(v)=w_+\frac{\bar{\gamma}_+^2\bar{v}_+}{\bar{\gamma}^2\bar{v}}=w_N\frac{v_w}{1-v_w^2}\frac{1-\mu(v_w,v)^2}{\mu(v_w,v)}
\end{align}
 with $\bar{v}_+=v_w$ and $\bar{v}=\mu(v_w,v)=\mu(\xi=v_w,v(\xi))$.
Therefore, the hydrodynamic backreaction force acting on the bubble wall alone is given by
\begin{align}\label{eq:BagEoSFbackWallDetonation}
\left.\frac{F_\mathrm{back}^\mathrm{hydro}}{A}\right|_\mathrm{wall}&=-(w_+-w_-)+\int_{v_-}^0\mathrm{d}v\frac{w(v)\gamma(v)^2}{\mu(v_w,v)},\nonumber\\
&=-(w_+-w_-)+\frac{v_-}{v_--v_w}w_+,
\end{align}
where $v_-\equiv v(\xi=v_w-\delta)$ in the limit of $\delta\to0$ is the fluid velocity just behind the wall in the plasma frame, and  $w_+\equiv w(\xi=v_w+\delta)$ and $w_-\equiv w(\xi=v_w-\delta)$ are the enthalpy values just in the front and back of the bubble wall in the limit of $\delta\to0$, respectively. Note that for detonation the enthalpy $w_+$ just in the front of the bubble wall is equal to the enthalpy $w_N$ at null infinity since there is no shockwave in the front of the wall to disturb the enthalpy flow. The sound-shell contribution from~\eqref{eq:BagEoSFshell} reads
\begin{align}
\frac{F_\mathrm{back}^\mathrm{hydro}}{A}\bigg|_\mathrm{shell}=-\frac14(w_--w_s),
\end{align}
where the enthalpy $w_s\equiv w(\xi=c_s)$ traced by sound speed is equal to the enthalpy $w_O$ at the bubble center. The total backreaction force therefore reads
\begin{align}\label{eq:BagEoSFbackDetonation}
\frac{F_\mathrm{back}^\mathrm{hydro}}{A}&=\frac{F_\mathrm{back}^\mathrm{hydro}}{A}\bigg|_\mathrm{shell}+\frac{F_\mathrm{back}^\mathrm{hydro}}{A}\bigg|_\mathrm{wall}\nonumber\\
&=\frac14w_s+\frac34w_-+\frac{v_w}{v_--v_w}w_+.
\end{align}
On the other hand, the driving force from~\eqref{eq:BagEoSFdrive} reads 
\begin{align}\label{eq:BagEoSFdriveDetonation}
\frac{F_\mathrm{drive}}{A}&=-\frac14(w_+-w_s)+\frac34\alpha_Nw_N
\end{align} 
with $w_N=w_+$ and $w_O=w_s$ for the detonation mode.

\subsection{Deflagration}

For the deflagration mode, the wall contribution is evaluated by
\begin{align}\label{eq:BagEoSFbackWallDeflagration}
\left.\frac{F_\mathrm{back}^\mathrm{hydro}}{A}\right|_\mathrm{wall}&=-(w_+-w_-)+\int_0^{v_+}\mathrm{d}v\frac{w(v)\gamma(v)^2}{\mu(v_w,v)}\nonumber\\
&=-(w_+-w_-)-\frac{v_+}{v_+-v_w}w_-
\end{align}
with $w(v)$ given by
\begin{align}
w(v)=w_-\frac{\bar{\gamma}_-^2\bar{v}_-}{\bar{\gamma}^2\bar{v}}=w_-\frac{v_w}{1-v_w^2}\frac{1-\mu(v_w,v)^2}{\mu(v_w,v)},
\end{align}
where $v_+\equiv v(\xi=v_w+\delta)$ in the limit of $\delta\to0$ is the fluid velocity just in the front of the wall in the plasma frame, while the shockfront contribution is evaluated by
\begin{align}
\left.\frac{F_\mathrm{back}^\mathrm{hydro}}{A}\right|_\mathrm{shock}&=-(w_R-w_L)+\int_{v(v_{sh})}^0\mathrm{d}v\frac{w(v)\gamma(v)^2}{\mu(v_{sh},v)}\nonumber\\
&=-(w_R-w_L)+\frac{v(v_{sh})}{v(v_{sh})-v_{sh}}w_N
\end{align}
with $w(v)$ given by
\begin{align}
w(v)=w_R\frac{\tilde{\gamma}_R^2\tilde{v}_R}{\tilde{\gamma}^2\tilde{v}}=w_N\frac{v_{sh}}{1-v_{sh}^2}\frac{1-\mu(v_{sh},v)^2}{\mu(v_{sh},v)},
\end{align}
where $v_{sh}$ is the velocity of the shockwave front, and $v(v_{sh})$ is the fluid velocity traced by $\xi=v_{sh}$ in the plasma frame, while $w_R\equiv w(\xi=v_{sh}+\delta)$ and $w_L\equiv w(\xi=v_{sh}-\delta)$ are the enthalpy values just in the front and back of the shockwave front in the limit of $\delta\to0$, respectively. The over-tilde symbol is introduced for the shockfront frame. The sound-shell contribution from~\eqref{eq:BagEoSFshell} reads
\begin{align}
\frac{F_\mathrm{back}^\mathrm{hydro}}{A}\bigg|_\mathrm{shell}=-\frac14(w_L-w_+).
\end{align}
Therefore, the total backreaction force reads
\begin{align}\label{eq:BagEoSFbackDeflagration}
\frac{F_\mathrm{back}^\mathrm{hydro}}{A}=\frac{F_\mathrm{back}^\mathrm{hydro}}{A}\bigg|_\mathrm{wall}+\frac{F_\mathrm{back}^\mathrm{hydro}}{A}\bigg|_\mathrm{shell}+\frac{F_\mathrm{back}^\mathrm{hydro}}{A}\bigg|_\mathrm{shock}.
\end{align}
On the other hand, the driving force from~\eqref{eq:BagEoSFdrive} reads
\begin{align}\label{eq:BagEoSFdriveDeflagration}
\frac{F_\mathrm{drive}}{A}&=-\frac14(w_R-w_-)+\frac34\alpha_Nw_N
\end{align} 
with $w_N=w_R$ and $w_O=w_-$ for the deflagration mode.

\subsection{Hybrid}

For the hybrid mode, the wall contribution is evaluated by
\begin{align}\label{eq:BagEoSFbackWallHybrid}
\left.\frac{F_\mathrm{back}^\mathrm{hydro}}{A}\right|_\mathrm{wall}&=-(w_+-w_-)+\int_{v_-}^{v_+}\mathrm{d}v\frac{w(v)\gamma(v)^2}{\mu(v_w,v)}\nonumber\\
&=-(w_+-w_-)+\frac{c_s(1-v_w^2)(v_--v_+)w_-}{(1-c_s^2)(v_--v_w)(v_w-v_+)}
\end{align}
with $w(v)$ given by
\begin{align}
w(v)=w_-\frac{\bar{\gamma}_-^2\bar{v}_-}{\bar{\gamma}^2\bar{v}}=w_-\frac{c_s}{1-c_s^2}\frac{1-\mu(v_w,v)^2}{\mu(v_w,v)},
\end{align}
and the shockfront contribution is evaluated by
\begin{align}
\left.\frac{F_\mathrm{back}^\mathrm{hydro}}{A}\right|_\mathrm{shock}&=-(w_R-w_L)+\int_{v(v_{sh})}^0\mathrm{d}v\frac{w(v)\gamma(v)^2}{\mu(v_{sh},v)}\nonumber\\
&=-(w_R-w_L)+\frac{v(v_{sh})}{v(v_{sh})-v_{sh}}w_N
\end{align}
with $w(v)$ given by
\begin{align}
w(v)=w_R\frac{\tilde{\gamma}_R^2\tilde{v}_R}{\tilde{\gamma}^2\tilde{v}}=w_N\frac{v_{sh}}{1-v_{sh}^2}\frac{1-\mu(v_{sh},v)^2}{\mu(v_{sh},v)}.
\end{align}
The sound-shell contribution from~\eqref{eq:BagEoSFshell} reads
\begin{align}
\frac{F_\mathrm{back}^\mathrm{hydro}}{A}\bigg|_\mathrm{shell}=-\frac14[(w_L-w_+)+(w_--w(c_s))].
\end{align}
The total  backreaction force therefore reads
\begin{align}\label{eq:BagEoSFbackHybrid}
\frac{F_\mathrm{back}^\mathrm{hydro}}{A}=\frac{F_\mathrm{back}^\mathrm{hydro}}{A}\bigg|_\mathrm{shell}+\left.\frac{F_\mathrm{back}^\mathrm{hydro}}{A}\right|_\mathrm{wall}+\left.\frac{F_\mathrm{back}^\mathrm{hydro}}{A}\right|_\mathrm{shock}.
\end{align}
On the other hand, the driving force from~\eqref{eq:BagEoSFdrive}
\begin{align}\label{eq:BagEoSFdriveHybrid}
\frac{F_\mathrm{drive}}{A}&=-\frac14(w_R-w_s)+\frac34\alpha_Nw_N
\end{align} 
with $w_N=w_R$ and $w_O=w_s$ for the hybrid mode.

\section{Numerical evaluation results}

\begin{figure*}
\centering
\includegraphics[width=0.48\textwidth]{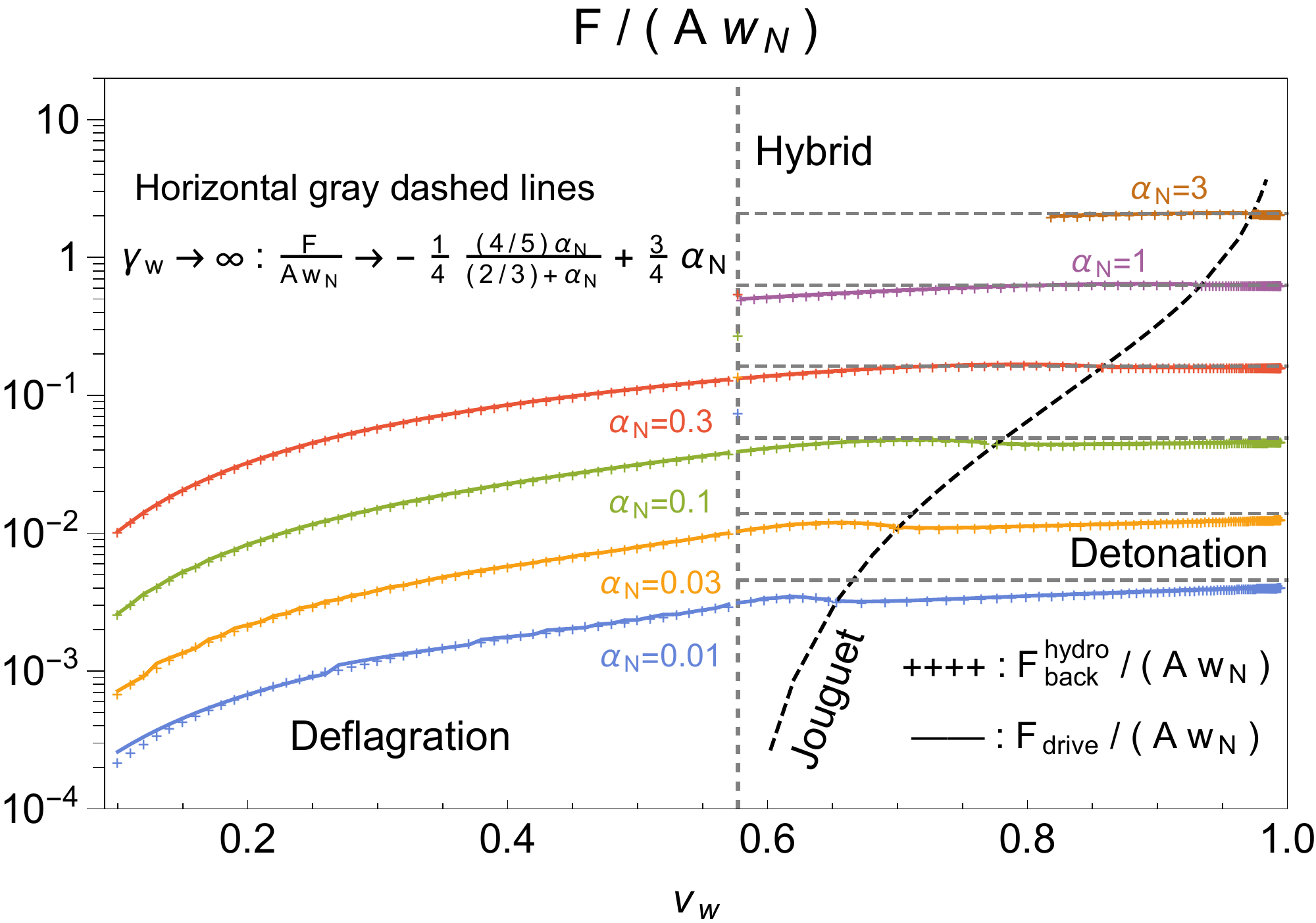}
\includegraphics[width=0.48\textwidth]{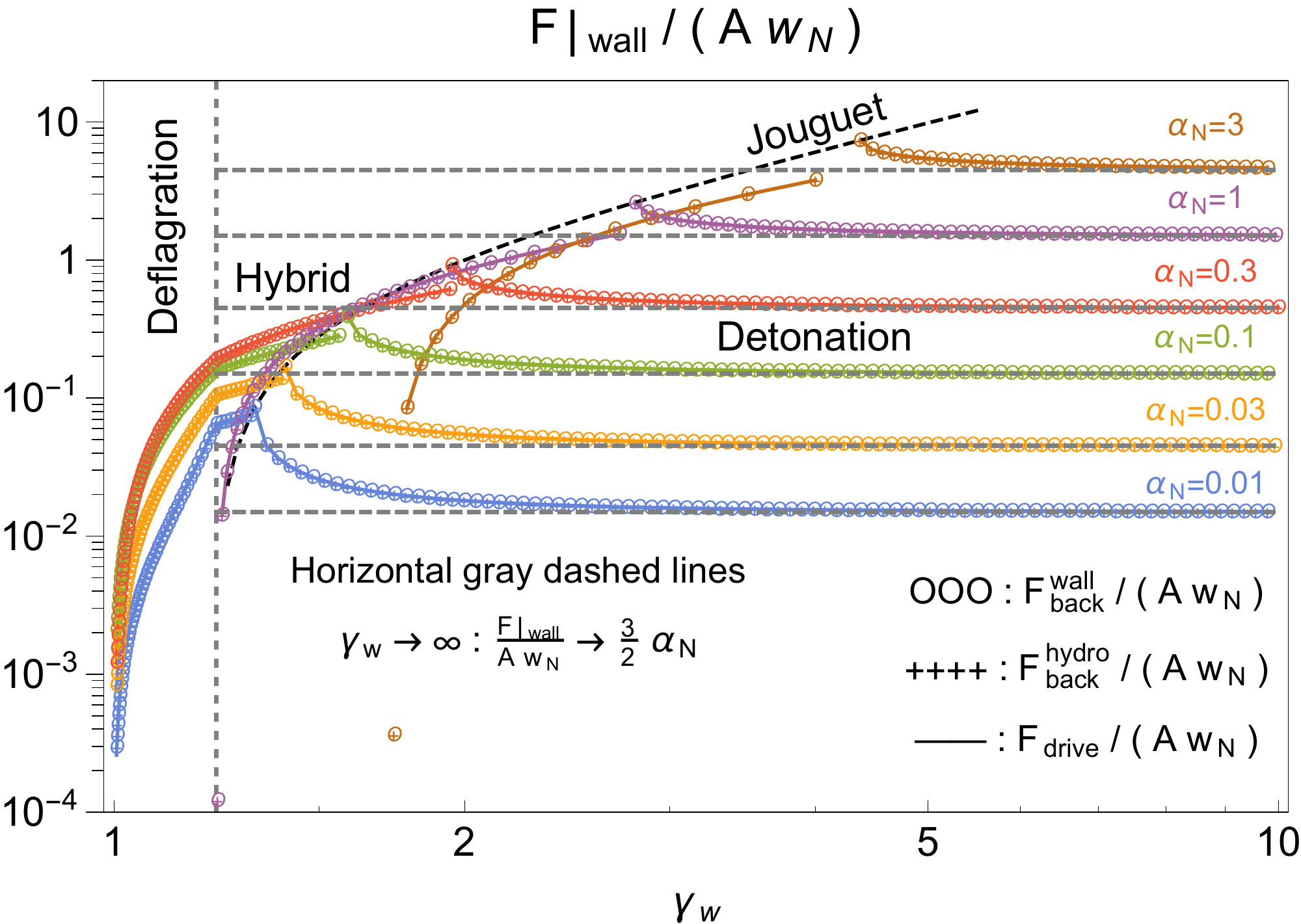}\\
\includegraphics[width=0.48\textwidth]{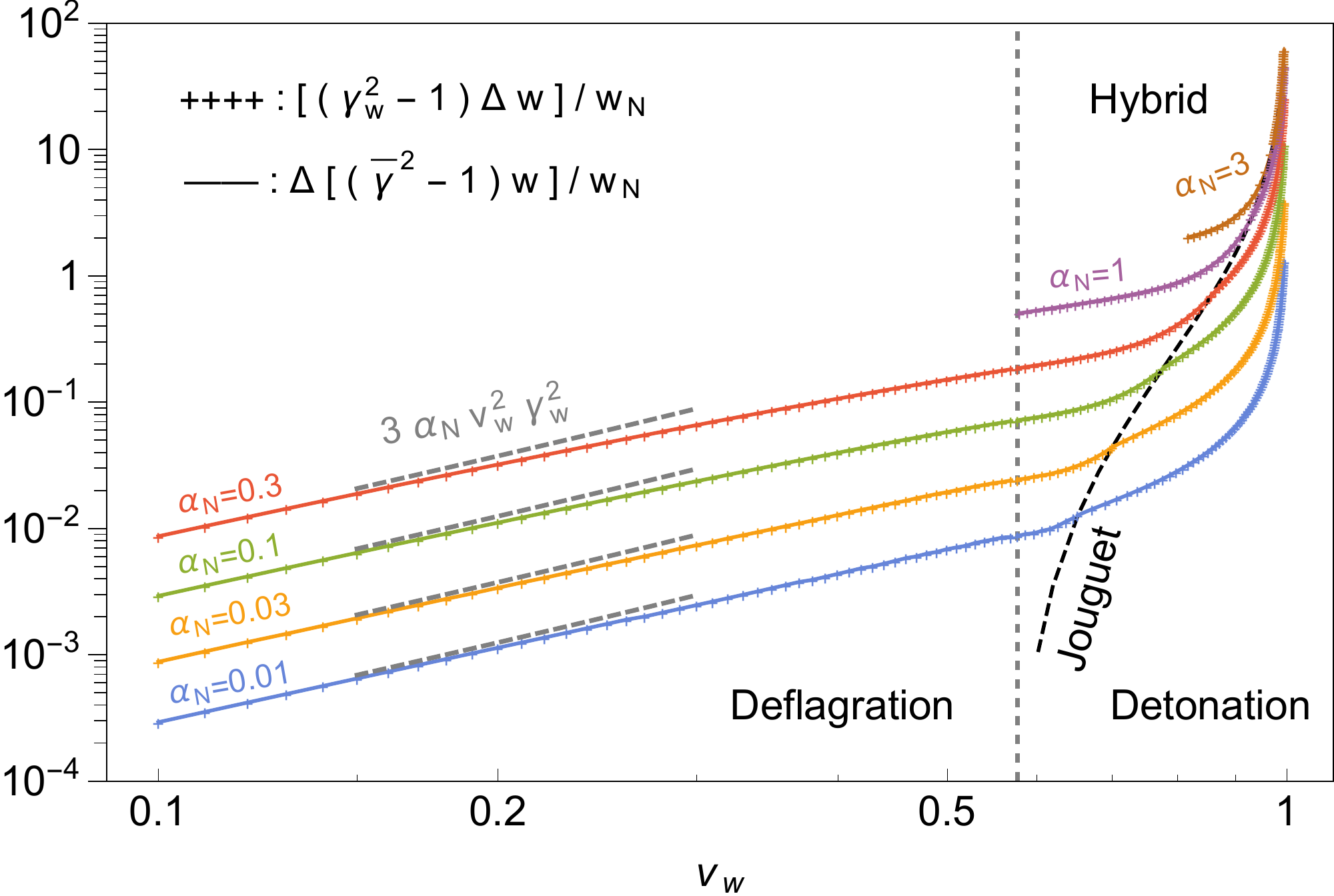}
\includegraphics[width=0.48\textwidth]{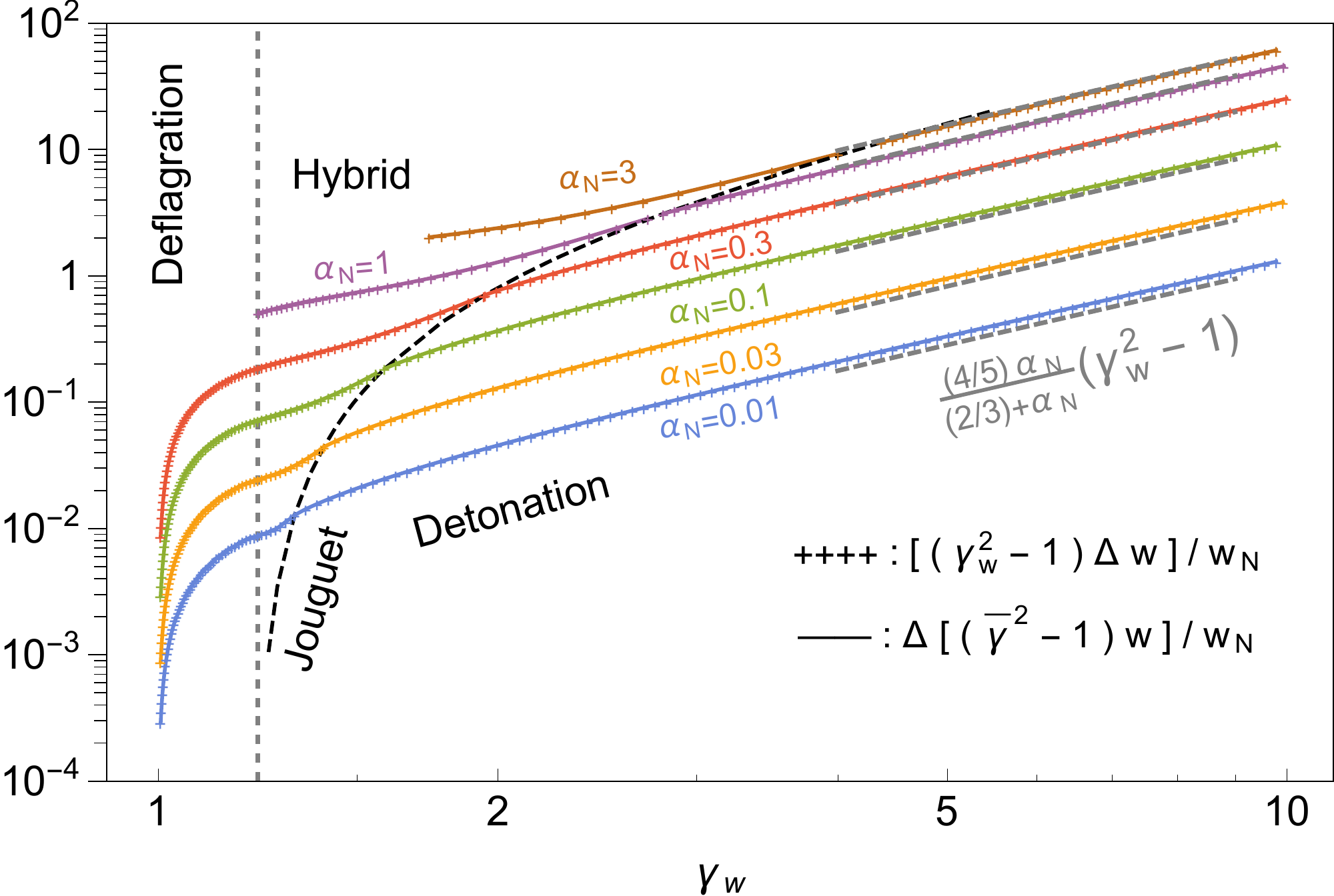}\\
\caption{In the first panel, our hydrodynamic backreaction force exactly matches the driving force. In the second panel, the wall contribution to our hydrodynamic backreaction force exactly matches the previous estimation $\Delta_\mathrm{wall}(w\bar{\gamma}^2\bar{v}^2)$, both of which are balanced by the driving force acting on the wall. In the last two panel, $\Delta(w\bar{\gamma}^2\bar{v}^2)=(\gamma_w^2-1)\Delta w$ are shown with respect to $v_w$ (left one) and $\gamma_w$ (right one). The asymptotic behaviours are indicated by the gray dashed lines.}
\label{fig:Fback}
\end{figure*}

The validity of our explicit hydrodynamic evaluation for the total (not only the friction force but also the thermal force) and full (not only the wall contribution but also the shell and shock contributions) backreaction force can be checked numerically as summarized in Fig.~\ref{fig:Fback} for some illustrative values of the strength factor $\alpha_N=0.01, 0.03, 0.1, 0.3, 1, 3$.

\subsection{The balance of forces}

In the first panel of Fig.~\ref{fig:Fback}, we compare our explicit hydrodynamics evaluation on the backreaction force~\eqref{eq:FbackHydro} (plus markers) with respect to the driving force~\eqref{eq:BagEoSFdrive} (solid curves), and find a perfect match numerically as expected by the balance of forces, which also serves as a consistency check for a non-runaway steady-state bubble expansion. In particular, the balance between the driving force~\eqref{eq:BagEoSFdriveDetonation} and backreaction force~\eqref{eq:BagEoSFbackDetonation} for the detonation expansion leads to an expression for the enthalpy just behind the bubble wall,
\begin{align}
w_-=\left[(1+\alpha_+)+\frac43\frac{v_-}{v_w-v_-}\right]w_+.
\end{align}
On the other hand, the first junction condition~\eqref{eq:1stjunction} also gives rise to
\begin{align}\label{eq:1stjuctionDetonation}
w_-=\frac{\bar{v}_+\bar{\gamma}_+^2}{\bar{v}_-\bar{\gamma}_-^2}w_+=\left(\frac{v_-}{v_w-v_-}+\frac{1}{1-v_wv_-}\right)w_+
\end{align}
with $\bar{v}_+=v_w$ and $\bar{v}_-=\mu(v_w,v_-)=(v_w-v_-)/(1-v_wv_-)$ for the detonation expansion. The equivalence of these two expressions results in a simple relation
\begin{align}
\alpha_+=\frac{v_-(3v_w^2-2v_-v_w-1)}{3(v_w-v_-)(1-v_-v_w)},
\end{align}
which is nothing but the minus branch of solutions~\cite{Espinosa:2010hh}
\begin{align}
\bar{v}_+=\frac{1}{1+\alpha_+}\left(X_+\pm\sqrt{X_-^2+\alpha_+^2+\frac23\alpha_+}\right)
\end{align}
with abbreviations $X_\pm\equiv\bar{v}_-/2\pm1/(6\bar{v}_-)$. Therefore, we can even analytically prove the balance of forces for the detonation expansion and hence the validity of our hydrodynamic evaluation on the backreaction force.

\subsection{Backreaction force at the wall interface}

In the second panel of Fig.~\ref{fig:Fback}, the wall contributions~\eqref{eq:BagEoSFbackWallDetonation}, \eqref{eq:BagEoSFbackWallDeflagration}, and~\eqref{eq:BagEoSFbackWallHybrid} to our hydrodynamic backreaction force~\eqref{eq:FbackHydro} (plus markers) exactly matches the previous proposal $\Delta_\mathrm{wall}(w\bar{\gamma}^2\bar{v}^2)$ (circle markers), both of which can be balanced by the driving force $\Delta_\mathrm{wall}(V_\mathrm{eff})$ at the wall interface (solid curves), where the results for the deflagration and hybrid modes are shown by their negative values. In particular, the match between~\eqref{eq:BagEoSFbackWallDetonation} and~\eqref{eq:balanceatwall} can be proved analytically for the detonation expansion through the equivalence
\begin{align}
\frac{F_\mathrm{back}^\mathrm{hydro}}{A}\bigg|_\mathrm{wall}=\frac{v_-v_w}{1-v_-v_w}w_+=\Delta_\mathrm{wall}(w\bar{\gamma}^2\bar{v}^2)
\end{align}
from $\bar{v}_+=v_w$ and $\bar{v}_-=\mu(v_w,v_-)=(v_w-v_-)/(1-v_wv_-)$ as well as~\eqref{eq:1stjuctionDetonation}. Therefore, our hydrodynamic evaluation on the backreaction force not only can balance the driving force but also reproduce the previous estimation $\Delta_\mathrm{wall}(w\bar{\gamma}^2\bar{v}^2)$ of the backreaction force acting on the wall.

It is worth noting that one cannot simply generalize the difference $\Delta_\mathrm{wall}(w\bar{\gamma}^2\bar{v}^2)$ taken near the wall into the difference $\Delta(w\bar{\gamma}^2\bar{v}^2)$ taken sufficiently distant away from the wall in order to evaluate the full backreaction force that balances the driving force. This is simply because that the pressure difference $-\Delta_\mathrm{wall}p=\Delta_\mathrm{wall}(w\bar{\gamma}^2\bar{v}^2)$ is only valid in the vicinity of the wall instead of anywhere away from the wall. In fact, in the third and last panels of Fig.~\ref{fig:Fback}, we have explicitly computed $\Delta(w\bar{\gamma}^2\bar{v}^2)$ (solid curves) and $\gamma_w^2v_w^2\Delta w$ (plus markers) with the differences taken outside the sound shell (or equivalently between $\xi=1$ and $\xi=0$). It is easy to see the equality 
$\Delta(w\bar{\gamma}^2\bar{v}^2)=\gamma_w^2v_w^2\Delta w$, but both of which do not resemble either the full (first panel) or wall (second panel) contributions to the total backreaction force. Therefore, although the difference $\Delta_\mathrm{wall}(w\bar{\gamma}^2\bar{v}^2)$ near the wall successfully capture the wall contribution of the backreaction force, its direct generalization to $\Delta(w\bar{\gamma}^2\bar{v}^2)$ does not recover the full backreaction force that should balance the driving force.  This is why we propose our hydrodynamic evaluation for the backreaction force.

\subsection{Ultra-relativistic behavior}\label{subsec:UR}

In the ultra-relativistic limit of the terminal wall velocity, both the driving force and backreaction force are fitted numerically to approach a $\gamma_w$-independent constant
\begin{align}
\frac{F_\mathrm{drive}}{Aw_N}=\frac{F_\mathrm{back}^\mathrm{hydro}}{Aw_N}\to-\frac14\frac{(4/5)\alpha_N}{(2/3)+\alpha_N}+\frac34\alpha_N
\end{align}
as shown by the asymptotic horizontal gray dashed lines in the first panel. As for the wall contribution shown in the second panel, their ultra-relativistic behaviors also approach a $\gamma_w$-independent constant
\begin{align}
\frac{F_\mathrm{back}^\mathrm{wall}}{Aw_N}=\left.\frac{F_\mathrm{back}^\mathrm{hydro}}{Aw_N}\right|_\mathrm{wall}=\left.\frac{F_\mathrm{drive}}{Aw_N}\right|_\mathrm{wall}\to\frac32\alpha_N
\end{align}
as shown by the horizontal gray dashed lines in the second panel, which can be proved analytically by noting that the wall-frame fluid velocity $\bar{v}_-$ just behind the bubble wall approaches $v_w-\frac32\alpha_N\gamma_w^{-2}$ and hence $v_-\to3\alpha_N/(2+3\alpha_N)$ at a large $\gamma_w$ limit. 
However, these constant ultra-relativistic limits do not mean that the backreaction force should admit no dependence on the bubble wall velocity since the hydrodynamics we are working with has presumed in the first place a non-runaway steady-state expansion with the driving force already balanced by the backreaction force. The true $\gamma_w$-dependence in the backreaction force (either in its thermal force or friction force components) should be obtained before reaching a steady-state, which will be reserved for future study.

In the last two panels, we naively evaluate $\Delta(w\bar{\gamma}^2\bar{v}^2)$ over the full profile of fluid motions, and find an exact match to $(\gamma_w^2-1)\Delta w$, both of which can be asymptotically fitted by
\begin{align}
(\gamma_w^2-1)\frac{\Delta w}{w_N}\to
\begin{cases}
3\alpha_N v_w^2\gamma_w^2, & \gamma_w\to1,\\
\frac{(4/5)\alpha_N}{(2/3)+\alpha_N}(\gamma_w^2-1), & \gamma_w\to\infty
\end{cases}
\end{align}
as shown by the gray dashed lines with respect to $v_w$ (left) and $\gamma_w$ (right) in the last two panels.  As we have already shown before that $\Delta(w\bar{\gamma}^2\bar{v}^2)=(\gamma_w^2-1)\Delta w$ does not reproduce the full backreaction force, their ultra-relativistic behavior certainly cannot be used to infer a $(\gamma_w^2-1)$-dependence for the full backreaction force.

\section{Conclusions}

The cosmological FOPT is a promising probe for many BSM new physics in the early Universe. The bubble expansion dynamics is crucial for the determination of the terminal wall velocity, which is the most uncertain parameter in determining the SGWB from the FOPT. The previous focus is primary on the friction force induced by the out-of-equilibrium effect, but overlooks the thermal force induced by the temperature saltation across the bubble wall (see, however, \cite{Laurent:2022jrs}). The combination of the thermal force and friction force gives rise to the total backreaction force that eventually balances the driving force for a non-runaway steady-state bubble expansion. 

It was recently found that this backreaction force could be depicted by the pressure difference $\Delta_\mathrm{wall}(\bar{\gamma}^2\bar{v}^2w)$ across the wall interface, which is derived from the junction condition of the energy-momentum tensor at the wall interface. However, the ultra-relativistic behavior of $\Delta_\mathrm{wall}(\bar{\gamma}^2\bar{v}^2w)$ does not admit a $(\gamma_w^2-1)$-dependence as naively expected in the previous literature.  Although a direct generalization of $\Delta_\mathrm{wall}(\bar{\gamma}^2\bar{v}^2w)$ into $\Delta(\bar{\gamma}^2\bar{v}^2w)$ could reproduce the ultra-relativistic $(\gamma_w^2-1)$-dependence, the difference $\Delta(\bar{\gamma}^2\bar{v}^2w)$ taken far outside and far inside of the wall is in fact not the pressure difference taken sufficiently distant away from the wall, which is actually given by our hydrodynamic evaluation on the backreaction force~\eqref{eq:FbackHydro} with its wall contribution~\eqref{eq:FbackWall} exactly reproduce the previous estimation $\Delta_\mathrm{wall}(\bar{\gamma}^2\bar{v}^2w)$.

\section{Discussions}

Similar to our hydrodynamic evaluation on the total backreaction force, the hydrodynamic expression~\eqref{eq:frichydro} for the friction force can also be split into a sound-shell part with a non-vanishing fluid velocity and discontinuous parts consisting of wall interface and shockwave front. It is easy to check that the sound-shell part is simply vanished due to the conserved entropy flow in the region where the fluid EoMs~\eqref{eq:dv} and~\eqref{eq:dw} are valid. However, the bubble-wall part does not necessarily vanish but requires an extra input from microphysics for the entropy density $s$ as a function of $v$ across the bubble wall since both $s$ and $v$ have experienced a jump across the bubble wall. This is also the case for the thermal force~\eqref{eq:themo}. Nevertheless, when adding up the thermal force and friction force, the terms involving with the entropy density have been canceled out with each other so that the total backreaction force can be purely evaluated from the hydrodynamics alone just as the driving force that it balances with. This is why we attribute the thermal force to the total backreaction force instead of a modification on the driving force as thought in the earlier studies~\cite{Espinosa:2010hh,Konstandin:2010dm}. The freedom to determine $s(v)$ at the wall interface from microphysics simply reflects the fact that the hydrodynamics alone cannot fully determine the form of the friction force, let alone its $\gamma_w$-dependence. We hope we could come back to this issue in future when the form of $s(v)$ can be determined externally at the wall interface.

Another crucial issue is the ultra-relativistic behavior, in which limit the local thermal equilibrium cannot be immediately established in the regions right after the ultra-relativistic bubble wall has just swept over. This failure of the local thermal equilibrium in the vicinity of the bubble wall leads to the ballistic approximation~\cite{Moore:1995si,BarrosoMancha:2020fay}.  Furthermore, the bag EoS also fails in the ultra-relativistic limit of the wall velocity, in which case the free energy density receives an extra mean-field contribution~\cite{Moore:1995si} in addition to the usual radiations in the bag EoS. However, our hydrodynamic expression~\eqref{eq:FbackHydro} for the total backreaction force is derived without absence of the out-of-equilibrium term and without assuming a bag EoS. Nevertheless,  the explicit hydrodynamic evaluations~\eqref{eq:BagEoSFbackDetonation}, \eqref{eq:BagEoSFbackDeflagration}, and~\eqref{eq:BagEoSFbackHybrid} on the total backreaction force for the bubble expansion of detonation, deflagration, or hybrid types have already assumed a bag EoS. Therefore, the ultra-relativistic behaviors we obtained in Section~\ref{subsec:UR} cannot be taken literally. In future works, we can adopt more general hydrodynamics to evaluate our total backreaction force by going beyond a simple bag EoS in the constant sound velocity model (also known as the $\nu$ model)~\cite{Giese:2020rtr,Giese:2020znk,Wang:2020nzm} and the varying sound velocity model~\cite{Wang:2022lyd}.

\begin{acknowledgments}
We thank for helpful discussions with Wen-Yuan Ai, Ryusuke Jinno, and Hongbao Zhang.
SJW is supported by the National Key Research and Development Program of China Grant  No.2021YFC2203004, No. 2020YFC2201501 and No.2021YFA0718304, 
the National Natural Science Foundation of China Grants 
%No. 11647601, No. 11821505, No. 11851302, No. 12047503, No. 11991052, No. 12075297, No. 12047558, and 
No. 12105344 and No. 12235019, 
the Key Research Program of the Chinese Academy of Sciences (CAS) Grant No. XDPB15, 
the Key Research Program of Frontier Sciences of CAS, 
and the Science Research Grants from the China Manned Space Project with No. CMS-CSST-2021-B01.
\end{acknowledgments}

%\iffalse
\appendix

\section{Hydrodynamics}\label{app:hydrodynamics}

In this appendix, we will attach the hydrodynamic details~\cite{Espinosa:2010hh} for a non-runaway steady-state bubble expansion with a bag EoS.  The starting point is a perfect fluid approximation for the total energy-momentum tensor $T^{\mu\nu}=wu^\mu u^\nu+p\eta^{\mu\nu}$ with $w\equiv e+p$ the total enthalpy. In the background plasma frame with a spherical coordinate system, the four velocity reads $u^\mu=\gamma(1,v,0,0)$ with $v=\mathrm{d}r/\mathrm{d}t$, while the fluid velocity $\bar{v}$ in a local wall frame moving along $z$ direction is computed by $\bar{v}\equiv\mu(v_w, v)$, where the abbreviation
\begin{align}
\mu(\xi,v(\xi))=\frac{\xi-v(\xi)}{1-\xi v(\xi)}
\end{align}
is introduced for the fluid velocity in a local comoving frame with velocity $\xi$. The self-similar coordinate $\xi\equiv r/t$ is introduced to trace a fluid element moving with a velocity $\xi$ in the background plasma frame so that $v(\xi)$ is the fluid velocity at $r=\xi t$ seen by an observer in the background plasma frame. In a local wall frame moving along $z$ direction, the fluid four-velocity is defined by $u^\mu=\bar{\gamma}(1,0,0,-\bar{v})$ to ensure a positive $\bar{v}$ and the corresponding $T_{\mu\nu}$ in the local wall frame reads
\begin{align}
T^{\mu\nu}=\left(\begin{array}{cccc}
w\bar{\gamma}^2-p & 0 & 0 & -w\bar{\gamma}^2\bar{v}  \\
 0 & p & 0 & 0\\
 0 & 0 & p & 0\\
 -w\bar{\gamma}^2\bar{v}  & 0 & 0 & w\bar{\gamma}^2\bar{v}^2+p
\end{array}\right).
\end{align}

\subsection{Junction conditions}

Before we solve the fluid equation of motion (EoM), one has to specify the junction conditions across the wall interface and shockwave front. The junction conditions at the wall interface are induced by the conservation equation of the total energy-momentum tensor in a local wall frame, $\partial_{\bar{t}}T^{\bar{t}\bar{t}}+\partial_{\bar{z}}T^{\bar{z}\bar{t}}=0$ and $\partial_{\bar{t}}T^{\bar{t}\bar{z}}+\partial_{\bar{z}}T^{\bar{z}\bar{z}}=0$, leading to
\begin{align}
w_-\bar{v}_-\bar{\gamma}_-^2&=w_+\bar{v}_+\bar{\gamma}_+^2,\label{eq:junction1}\\
w_-\bar{v}_-^2\bar{\gamma}_-^2+p_-&=w_+\bar{v}_+^2\bar{\gamma}_+^2+p_+,\label{eq:junction2}
\end{align}
where $w_\pm=e_\pm+p_\pm$ and $\bar{v}_\pm$ are the total enthalpy and fluid velocities just in the front and back of the bubble wall in the bubble-wall frame, respectively, and $\bar{\gamma}_\pm=(1-\bar{v}_\pm^2)^{-1/2}$ are the corresponding Lorentz factors. The junction conditions at the shockwave front are obtained similarly with the subscripts ``$\pm$'' replaced by $R/L$ representing the front and back of the shockwave front.

The junction conditions~\eqref{eq:junction1} and~\eqref{eq:junction2} can be put in use when combining with a bag equation-of-state (EoS). If no particle during the penetration across the bubble wall acquires a mass comparable to the background temperature, then this scalar-plasma system can be approximately treated as a simple collection of the vacuum potential energy density $V_0$ and the ideal thermal gas, namely
\begin{align}\label{eq:bagEoS}
e_\pm=a_\pm T^4+V_0^\pm, \quad p_\pm=\frac13a_\pm T^4-V_0^\pm,
\end{align}
where $a_\pm\equiv(\pi^2/30)g_\mathrm{eff}^\pm$ are the effective number of relativistic degrees of freedom in the false and true vacua, respectively. With this bag EoS approximation, the junction conditions can be solved as
\begin{align}
\bar{v}_+\bar{v}_-&=\frac{p_+-p_-}{e_+-e_-}=\frac{1-(1-3\alpha_+)r}{3-3(1+\alpha_+)r},\\
\bar{v}_+/\bar{v}_-&=\frac{e_-+p_+}{e_++p_-}=\frac{3+(1-3\alpha_+)r}{1+3(1+\alpha_+)r},
\end{align}
with the abbreviations 
\begin{align}
\alpha_+=\frac{\Delta V_0}{a_+T_+^4}=\frac{4\Delta V_0}{3w_+}, \quad r=\frac{w_+}{w_-}=\frac{a_+ T_+^4}{a_-T_-^4}.
\end{align}
$\bar{v}_\pm$ can now be re-expressed as
\begin{align}
\bar{v}_+(\alpha_+,r)&=\sqrt{\frac{1-(1-3\alpha_+)r}{3-3(1+\alpha_+)r}\cdot\frac{3+(1-3\alpha_+)r}{1+3(1+\alpha_+)r}},\\
\bar{v}_-(\alpha_+,r)&=\sqrt{\left.\frac{1-(1-3\alpha_+)r}{3-3(1+\alpha_+)r}\right/\frac{3+(1-3\alpha_+)r}{1+3(1+\alpha_+)r}}.
\end{align}

\subsection{Fluid equation of motion}

The fluid EoM is derived from projecting the conservation equation of the total energy-momentum tensor parallel along and perpendicular to the fluid flow directions defined by $u^\mu=\gamma(v)(1,v,0,0)$ and $\tilde{u}^\mu=\gamma(v)(v,1,0,0)$ in the background plasma frame with a spherical coordinate system, respectively, namely $u_\nu\nabla_\mu T^{\mu\nu}=0$ and $\tilde{u}^\nu\nabla^\mu T_{\mu\nu}=0$, which, after using the relations $u_\mu u^\mu=-1$, $\tilde{u}_\mu\tilde{u}^\mu=1$, $u_\nu\nabla_\mu u^\nu=0$, $\tilde{u}_\mu u^\mu=0$, become
\begin{align}
u^\mu\nabla_\mu e&=-w\nabla_\mu u^\mu,\\
\tilde{u}^\mu\nabla_\mu p&=-w\tilde{u}^\nu u^\mu\nabla_\mu u_\nu.
\end{align}
After written in the plasma frame with a spherical coordinate system for $\nabla_\mu u^\mu=(2v/\xi)(\gamma/t)+(\gamma^3/t)(1-\xi v)\partial_\xi v$, the projected conservation equations becomes
\begin{align}
(\xi-v)\frac{\partial_\xi e}{w}&=2\frac{v}{\xi}+\gamma^2(1-\xi v)\partial_\xi v,\label{eq:de}\\
(1-\xi v)\frac{\partial_\xi p}{w}&=\gamma^2(\xi-v)\partial_\xi v,\label{eq:dp}
\end{align}
which can be rearranged by division and summation as
\begin{align}
2\frac{v}{\xi}&=\gamma^2(1-\xi v)\left(\frac{\mu(\xi,v)^2}{c_s^2}-1\right)\frac{\mathrm{d}v}{\mathrm{d}\xi},\label{eq:dv}\\
\frac{\mathrm{d}w}{\mathrm{d}\xi}&=w\gamma^2\mu(\xi,v)\left(\frac{1}{c_s^2}+1\right)\frac{\mathrm{d}v}{\mathrm{d}\xi}.\label{eq:dw}
\end{align}
with the sound speed formally defined by  $c_s^2=\partial_\xi p/\partial_\xi e$. For a bag EoS, one simply takes $c_s^2=1/3$.

For different expansion modes of detonation, deflagration and hybrid types,  the fluid equation~\eqref{eq:dv} for the velocity profile can be solved with appropriate junction conditions at the wall interface or/and shockwave front, which in turn solves the enthalpy profile by~\eqref{eq:dw} as
\begin{align}\label{eq:wxi}
w(\xi)=w(\xi_0)\exp\left[4\int_{\xi_0}^{\xi}\gamma[v(\zeta)]^2\mu[\zeta,v(\zeta)]v'(\zeta)\mathrm{d}\zeta\right]
\end{align}
for some initial point $\xi_0$ on the non-vanishing part of the fluid velocity profile. The fluid velocity profile also solves the temperature profile $\partial_\xi\ln T=\gamma^2\mu\partial_\xi v$ from the thermodynamic relation $w=T\frac{\partial p}{\partial T}$ and Eq.~\eqref{eq:dp} as
\begin{align}\label{eq:Txi}
T(\xi)=T(\xi_0)\exp\left[\int_{\xi_0}^{\xi}\gamma[v(\zeta)]^2\mu[\zeta,v(\zeta)]v'(\zeta)\mathrm{d}\zeta\right].
\end{align}
As noted shortly below that both profiles of the fluid velocity and enthalpy depend only on $v_w$ and $\alpha_+$, but the temperature profile admits extra dependence on the ratio $a_+/a_-$ from the junction conditions.

\begin{figure*}
\centering
\includegraphics[width=0.3\textwidth]{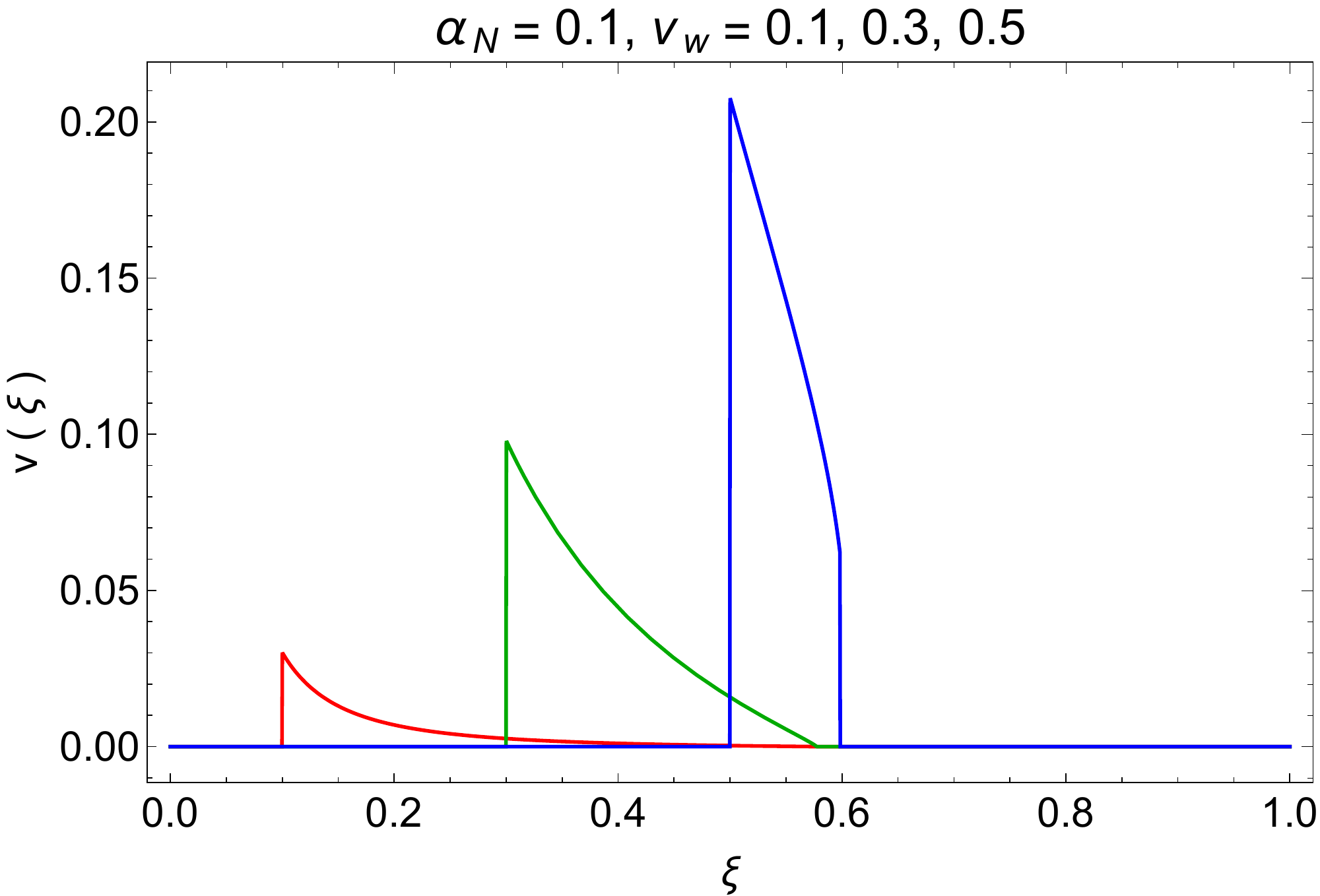}
\includegraphics[width=0.3\textwidth]{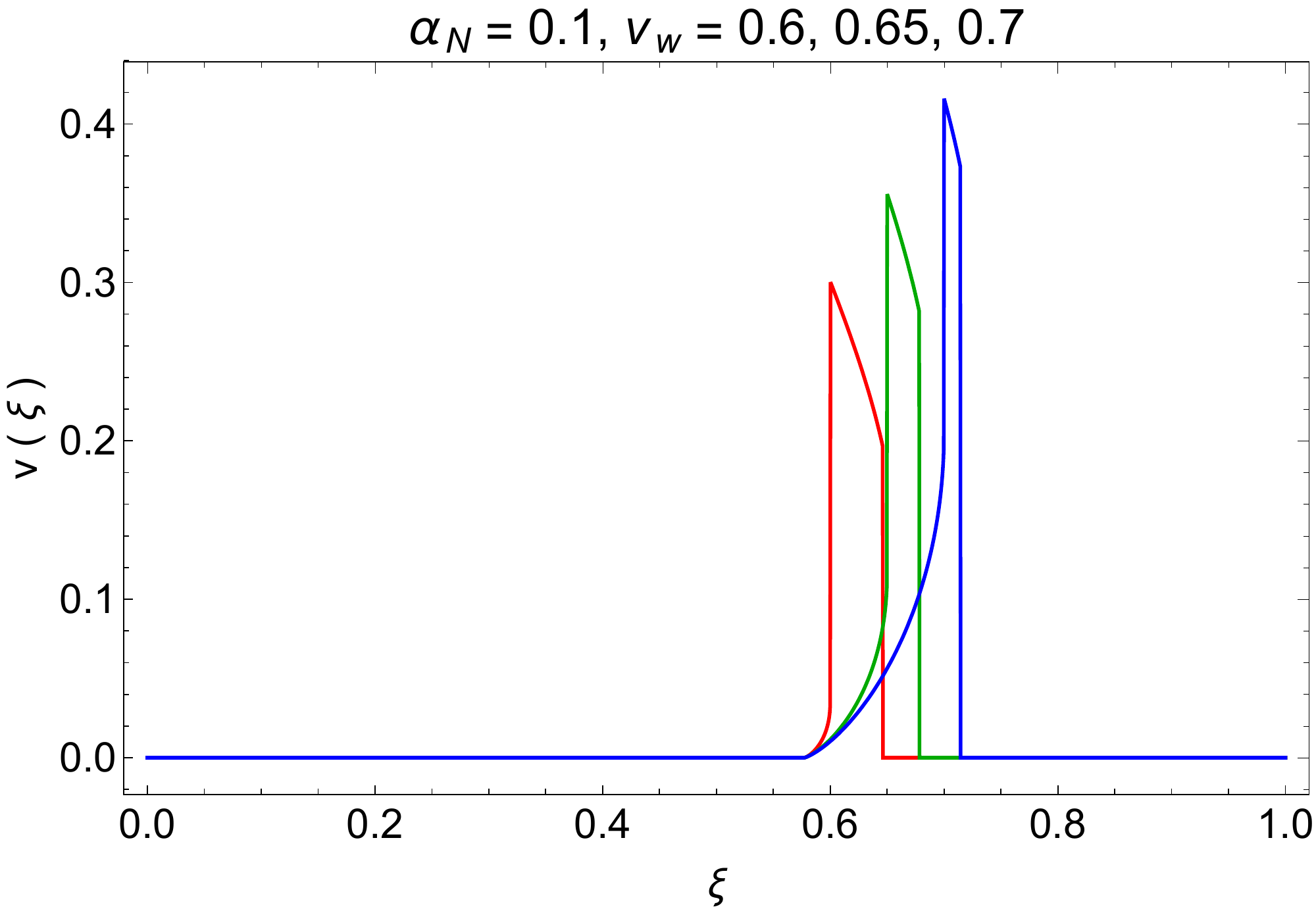}
\includegraphics[width=0.3\textwidth]{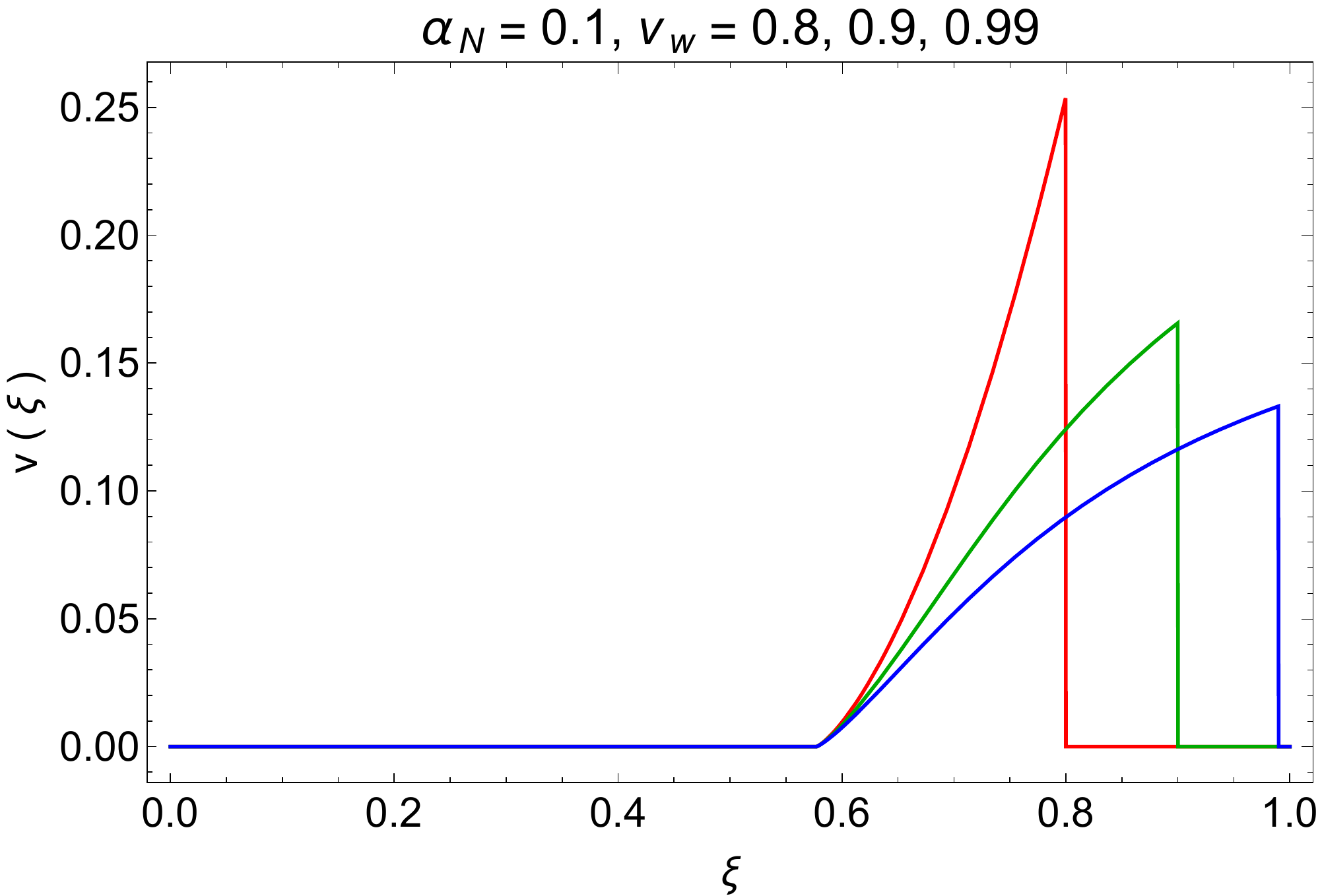}\\
\includegraphics[width=0.3\textwidth]{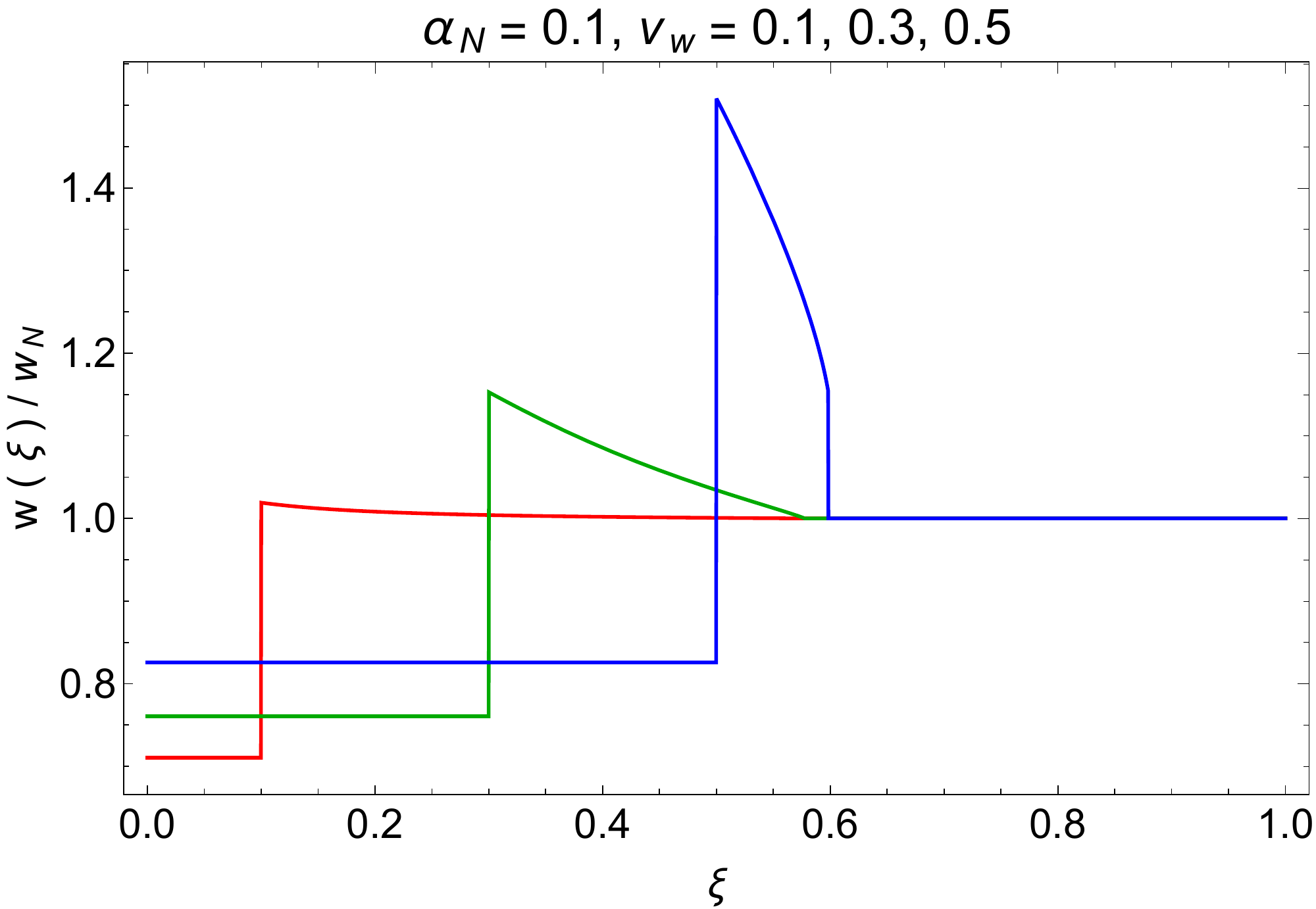}
\includegraphics[width=0.3\textwidth]{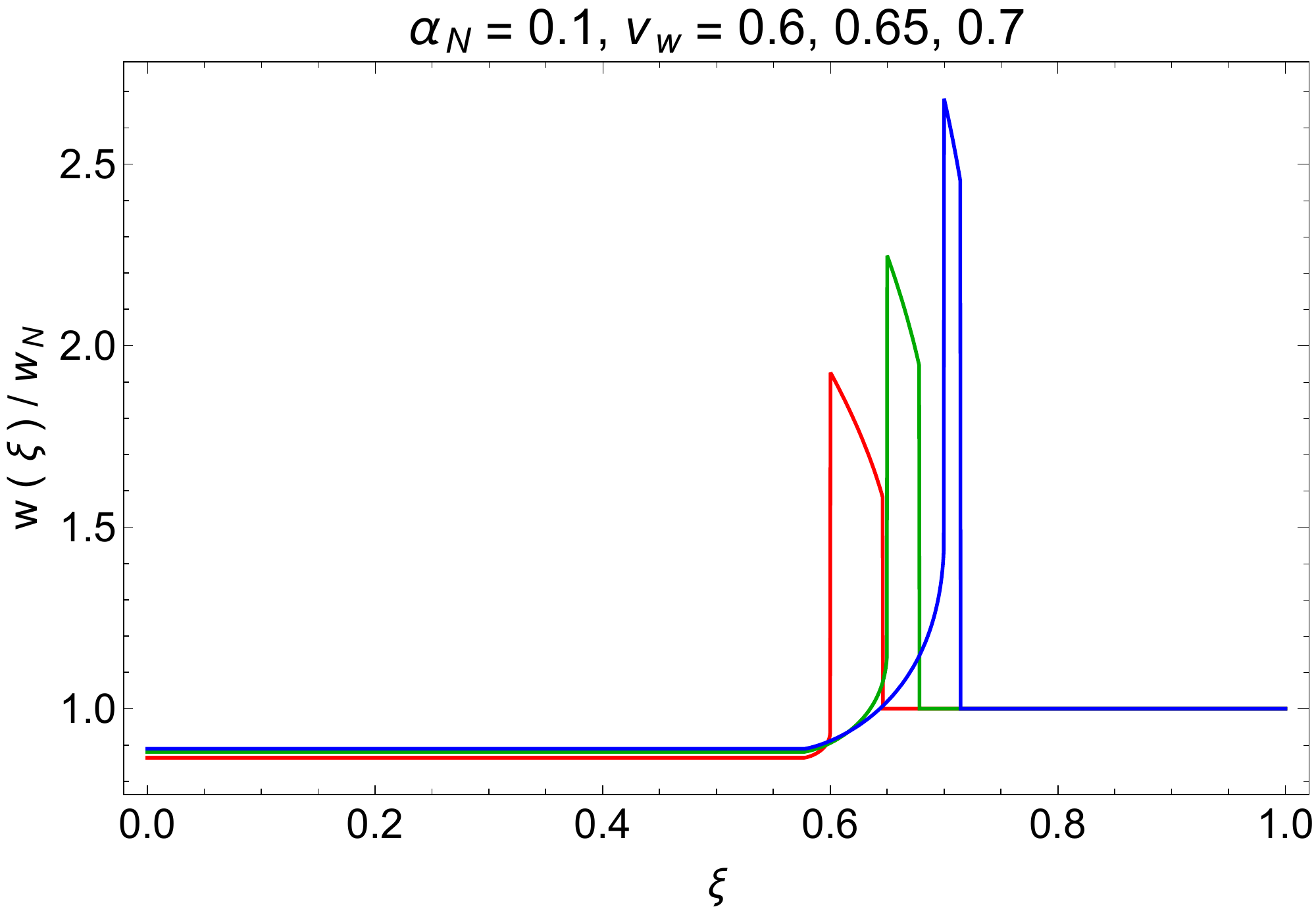}
\includegraphics[width=0.3\textwidth]{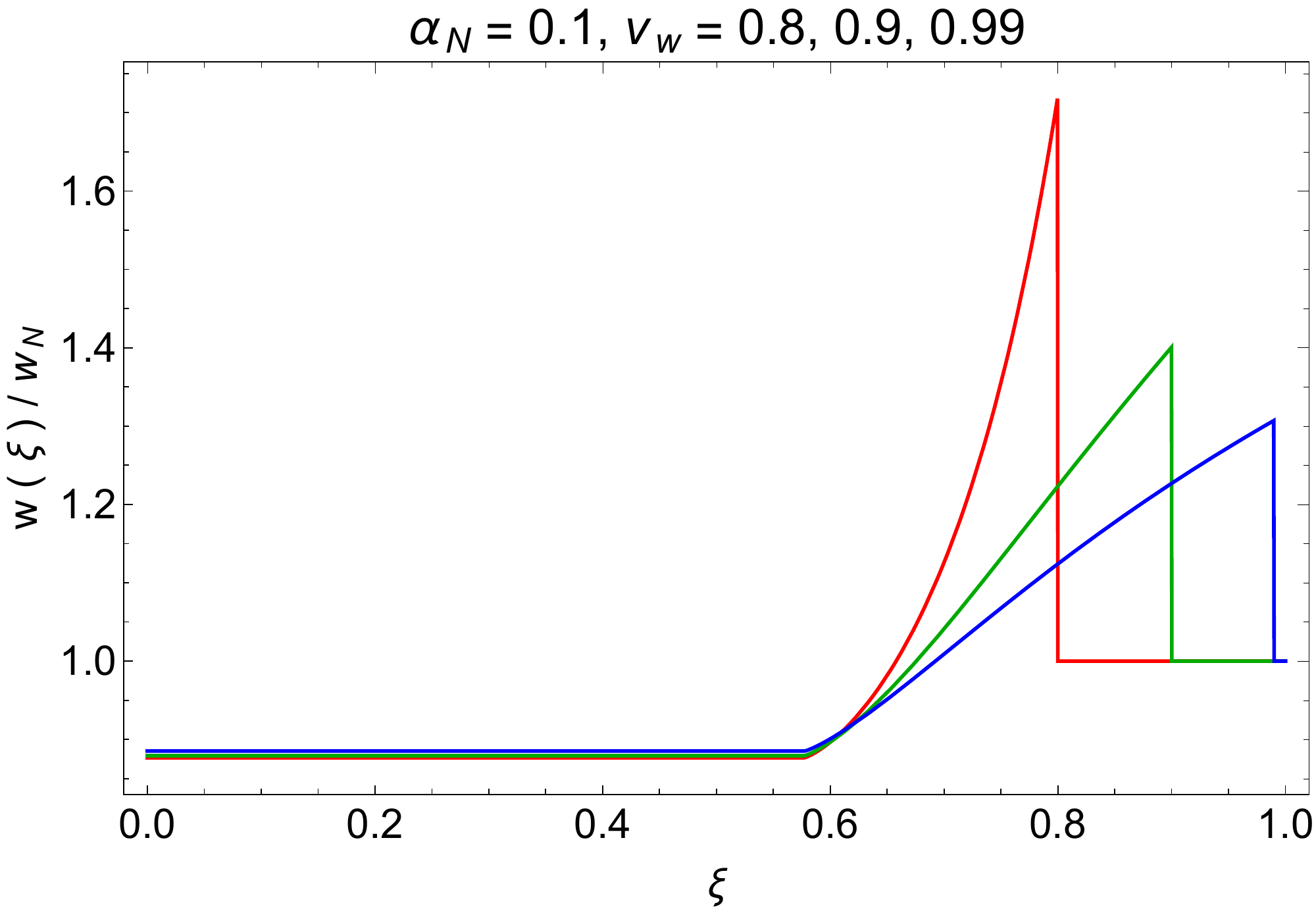}\\
\includegraphics[width=0.3\textwidth]{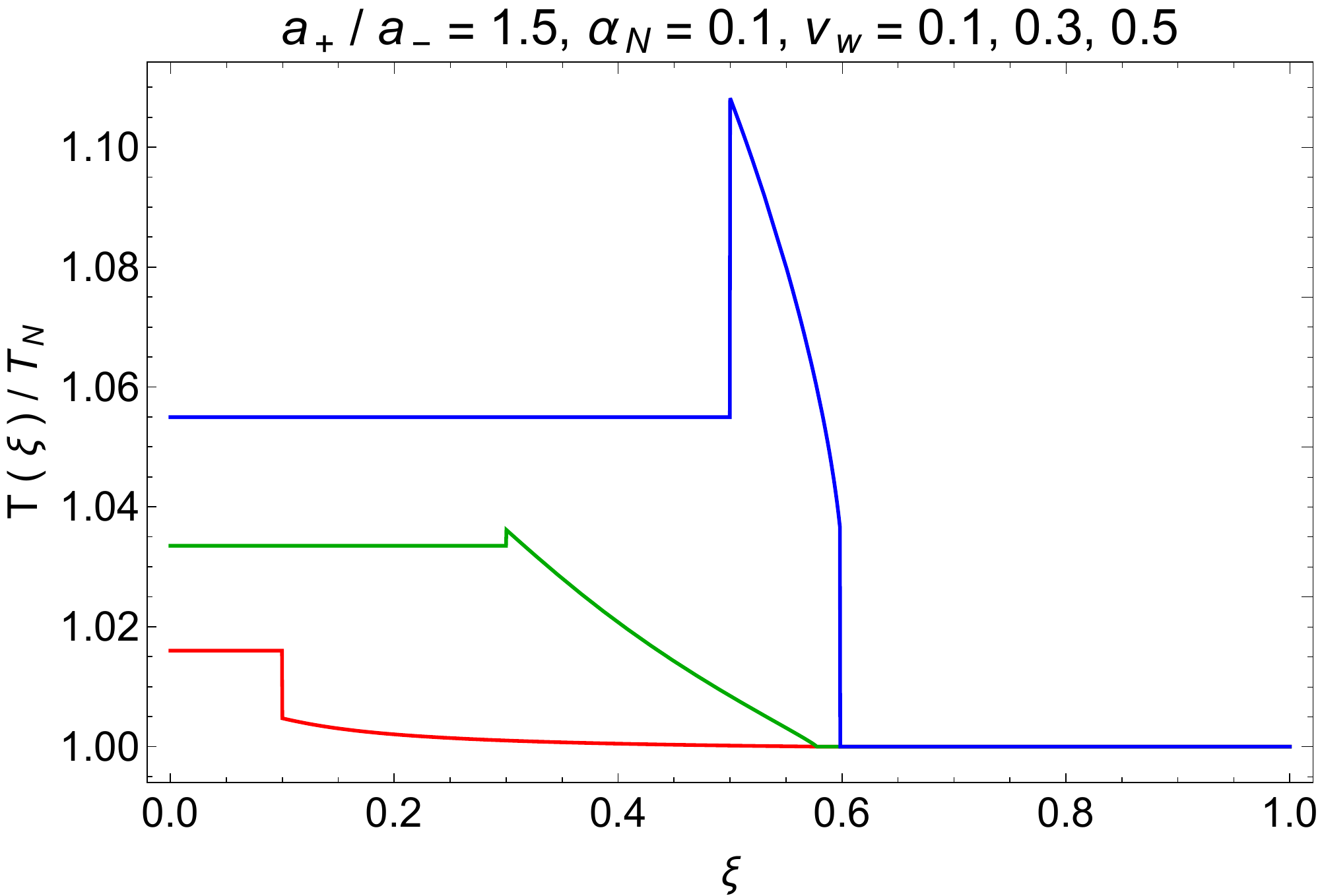}
\includegraphics[width=0.3\textwidth]{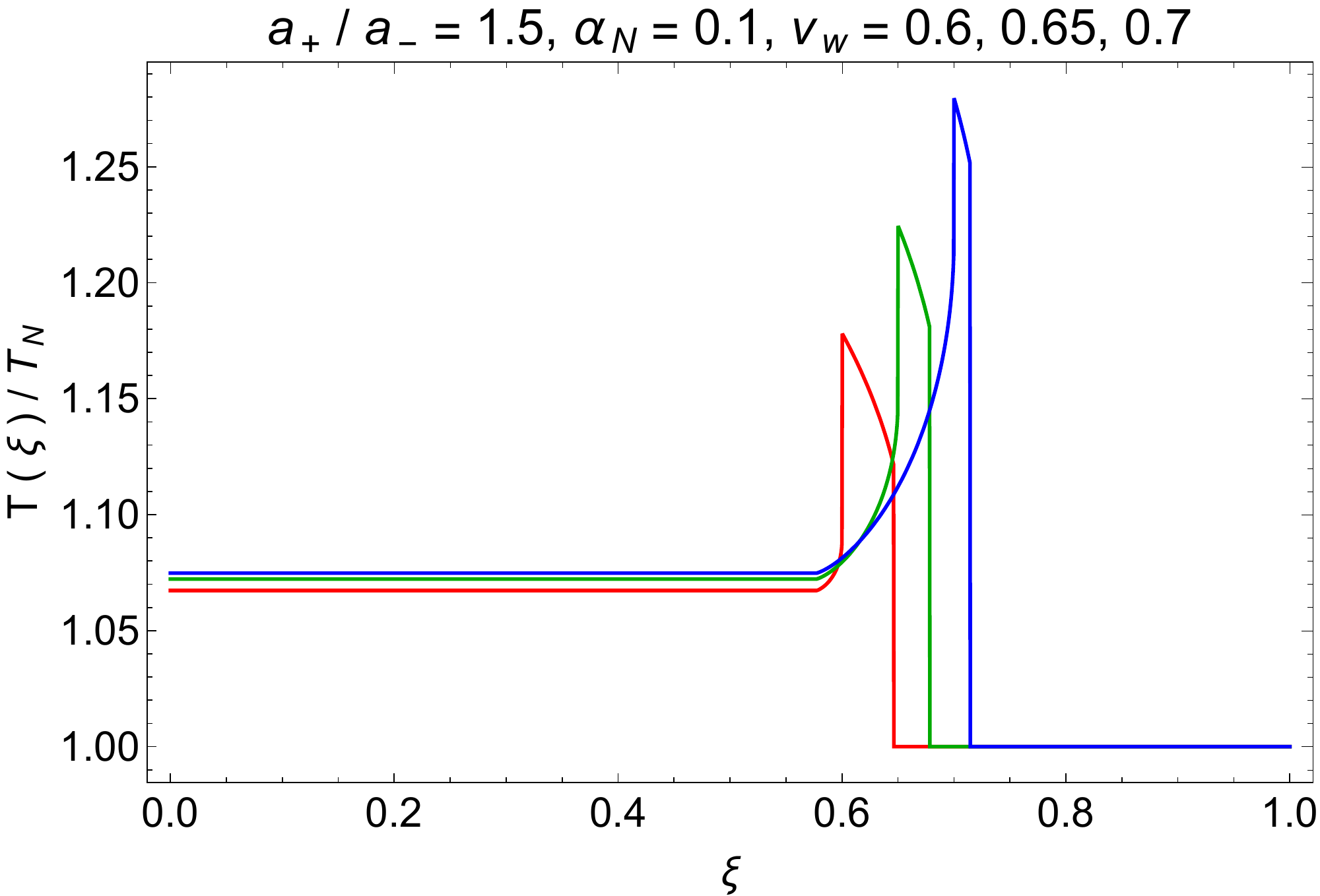}
\includegraphics[width=0.3\textwidth]{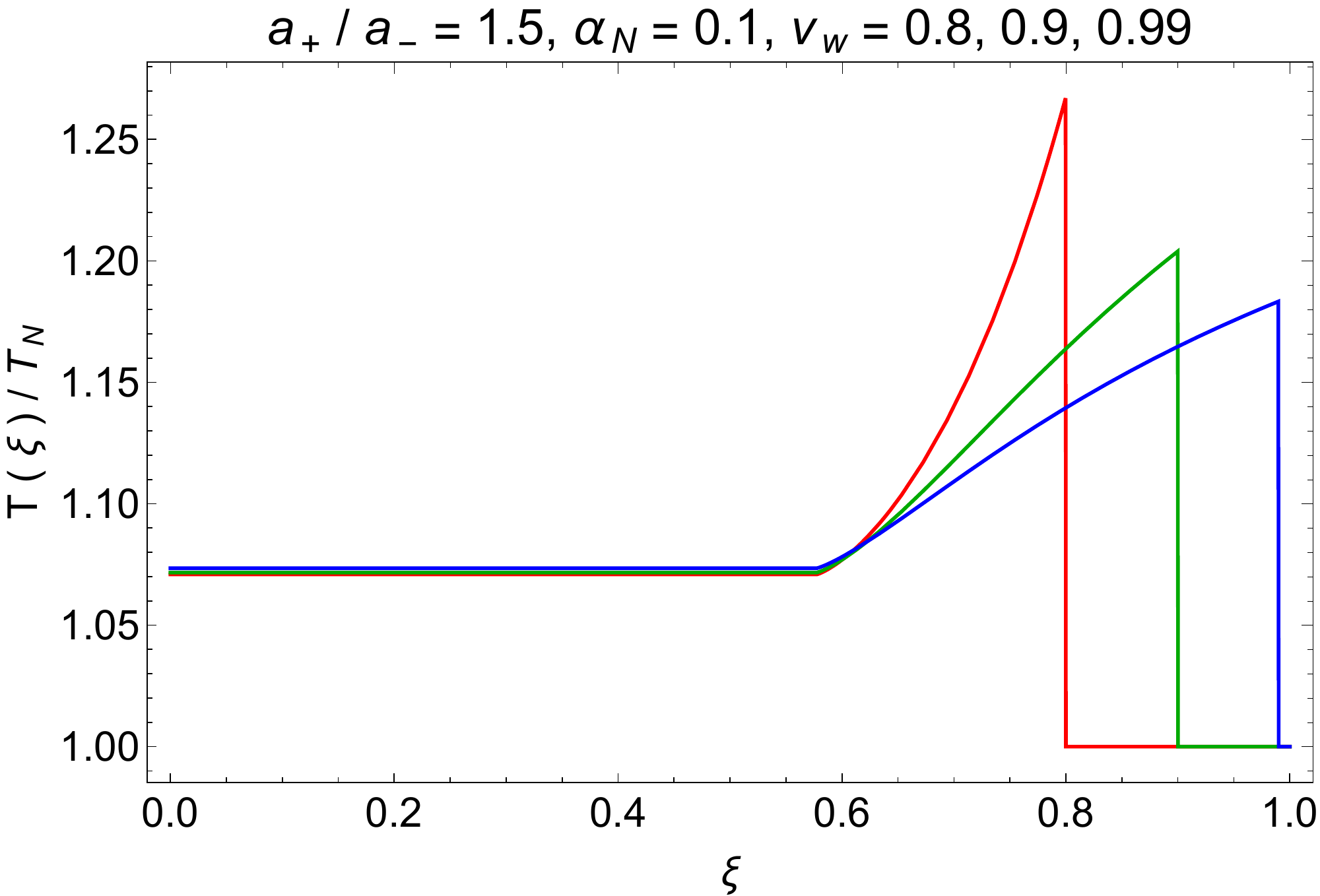}\\
\includegraphics[width=0.3\textwidth]{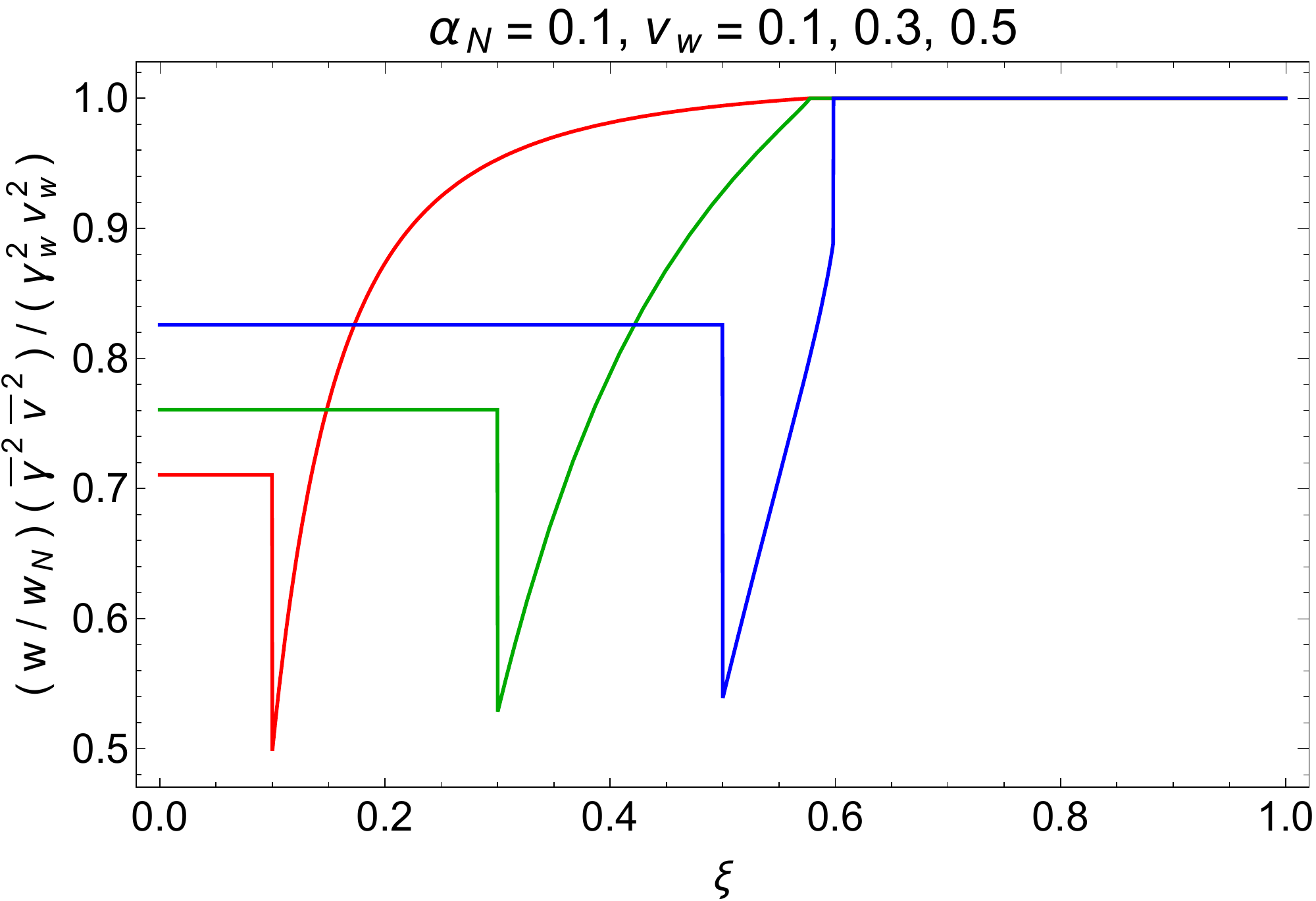}
\includegraphics[width=0.3\textwidth]{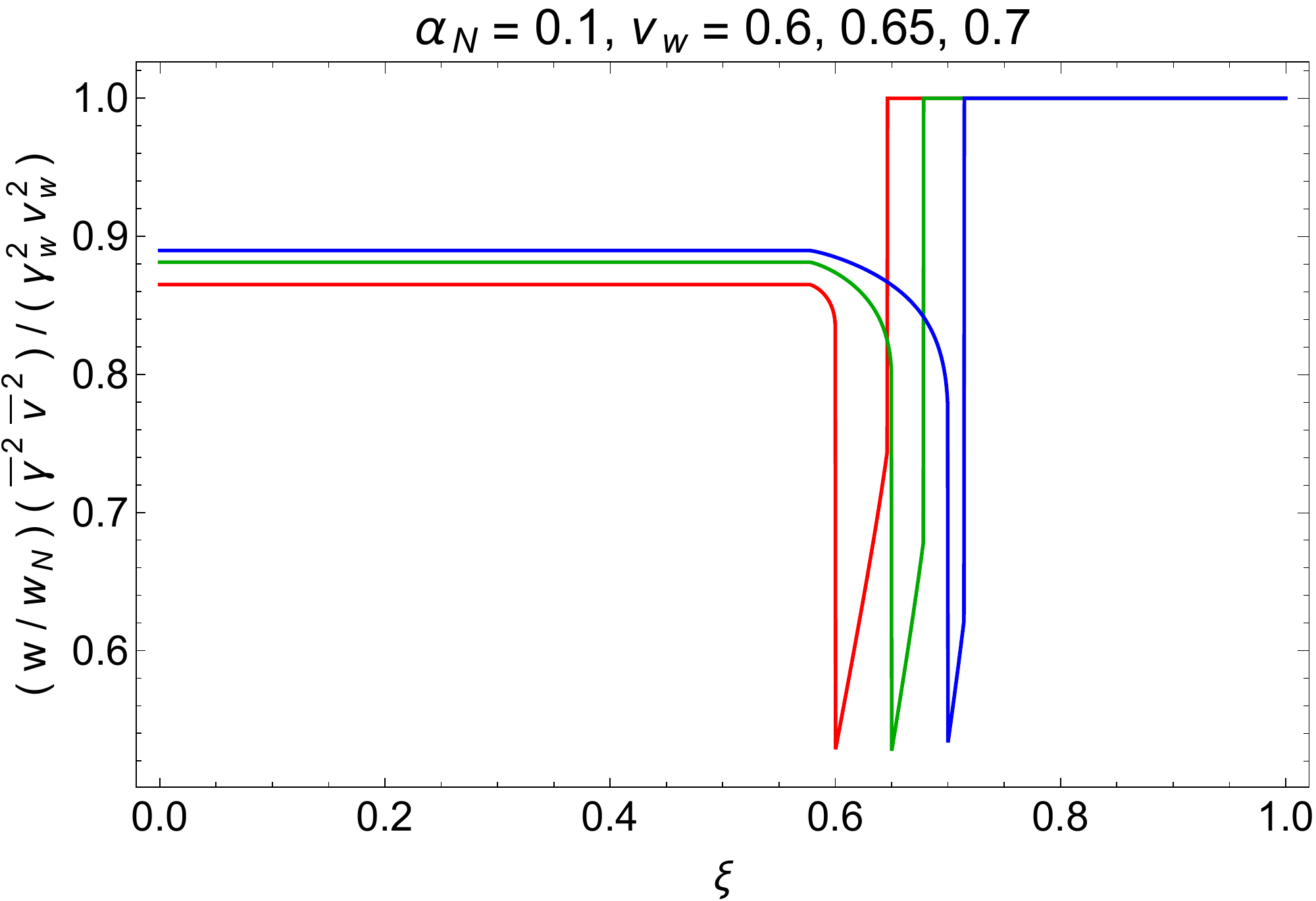}
\includegraphics[width=0.3\textwidth]{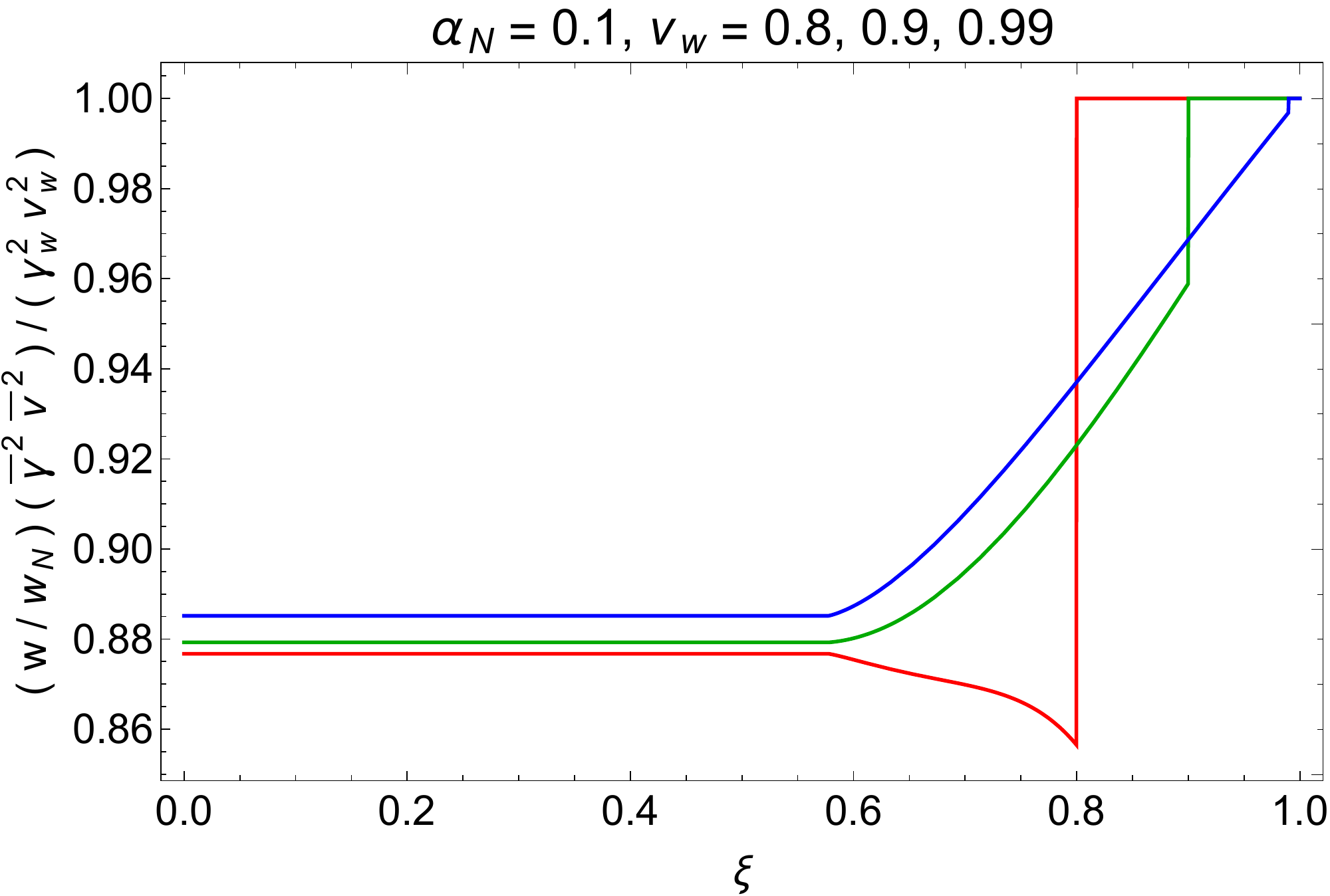}\\
\caption{The profiles of fluid velocity (first row), enthalpy $w/w_N$ (second row), temperature $T/T_N$ (third row), and $(w\bar{v}^2\bar{\gamma}^2)/(w_Nv_w^2\gamma_w^2)$ (fourth row) for expansion modes of detonation (right column), deflagration (left column), and hybrid (middle column) types.}
\label{fig:expansionmodes}
\end{figure*}

We summarize in Fig.~\ref{fig:expansionmodes} the profiles of the fluid velocity (first row), enthalpy $w/w_N$ (second row), and temperature $T/T_N$ (third row) as well as $(w\bar{\gamma}^2\bar{v}^2)/(w_N\gamma_w^2v_w^2)$ (fourth row) for the expansion modes of detonation (right column), deflagration (left column) and hybrid (middle column) types. In all cases, the asymptotic strength factor is fixed at $\alpha_N=0.1$ by
\begin{align}
\alpha_N=\alpha_+\frac{a_+T_+^4}{a_NT_N^4}=\alpha_+\frac{w(\xi^+=v_w)}{w_N},
\end{align}
and the temperature profile requires an extra parameter  $a_+/a_-=1.5$ for illustration. Some illustrative terminal wall velocity are chosen for the detonation mode with $v_w=0.8,0.9,0.99$, the deflagration mode with $v_w=0.1,0.2,0.3$, and the hybrid mode with $v_w=0.6,0.65,0.7$ in red, green, and blue colors in each case.

\subsubsection{Detonation}

The physically viable detonation mode ($\bar{v}_+>\bar{v}_-$) is of the weak type ($\bar{v}_->c_s$), which can be realized by a supersonic wall velocity larger than the Jouguet velocity $v_J$,
\begin{align}
v_w>v_J\equiv\frac{\sqrt{\alpha_+(2+3\alpha_+)}+1}{\sqrt{3}(1+\alpha_+)}>\frac{1}{\sqrt{3}}=c_s,
\end{align}
as we will see shortly below. For detonation mode, there is no shockwave but only rarefraction wave behind the bubble wall so that the fluid velocity in front of the bubble wall is zero in the background plasma frame, $v_+=\mu(v_w,\bar{v}_+)=0$, leading to  $v_w=\bar{v}_+(\alpha_+,r)$, which in turn determines $r(\alpha_+, v_w)$ as a function of $\alpha_+$ and $v_w$, so does $\bar{v}_\pm(\alpha_+,r(\alpha_+,v_w))\equiv\bar{v}_\pm(\alpha_+,v_w)$. Note that the previous requirement $v_w>v_J$ for the detonation realization can be directly solved from  $c_s<\bar{v}_-(\alpha_+,v_w)<v_w(\equiv\bar{v}_+)$. Then the fluid velocity just behind the bubble wall in the background plasma frame $v_-=\mu(v_w,\bar{v}_-)$ can be known from $\bar{v}_-(\alpha_+,v_w)$. Therefore, the fluid velocity profile $v(\xi)$ can be solved from Eq.~\eqref{eq:dv} with the initial condition chosen at $(v_w, v_-)$. 

The enthalpy profile can be obtained from Eq.~\eqref{eq:wxi} by specifying $\xi_0=v_w$ and $w(\xi_0)=w_-$, where $w_-$ is the enthalpy just behind the bubble wall given by
\begin{align}
w_-=w_+\frac{\bar{v}_+\bar{\gamma}_+^2}{\bar{v}_-\bar{\gamma}_-^2}=w_N\frac{v_w}{1-v_w^2}\frac{1-\bar{v}_-^2}{\bar{v}_-}
\end{align}
from the first junction condition~\eqref{eq:junction1} with $w_+\equiv w_N$, $\bar{v}_+=v_w$, and  $\bar{v}_-(\alpha_+,v_w)$. The temperature profile can also be obtained from Eq.~\eqref{eq:Txi} by specifying $\xi_0=v_w$ and $T(\xi_0)=T_-$, where $T_-$ is the temperature just behind the bubble wall given by
\begin{align}
T_-=\left(\frac{a_+/a_-}{r(\alpha_+,v_w)}\right)^{1/4}T_N
\end{align}
from the definition of $r=(a_-T_-^4)/(a_+T_+^4)$ with $T_+\equiv T_N$. Note that $w_+\equiv w_N$ and $T_+\equiv T_N$ are the same as its asymptotic values due to the absence of the shockwave to disturb the vacuum energy injection into the fluid in the front of the bubble wall. 

\subsubsection{Deflagration}

The physically viable deflagration mode ($\bar{v}_+<\bar{v}_-$) is also of the weak type ($\bar{v}_-<c_s$), which can be realized by a subsonic wall velocity $c_s>\bar{v}_-=v_w$ (see below) with a compressive shockwave in the front of the bubble wall. The fluid velocity behind the bubble wall is at rest in the background plasma frame, $v_-=\mu(v_w,\bar{v}_-)=0$, leading to $v_w=\bar{v}_-(\alpha_+,r)$,  which in turn determines $r(\alpha_+,v_w)$ as a function of $\alpha_+$ and $v_w$, so does $\bar{v}_\pm(\alpha_+, r(\alpha_+,v_w))\equiv\bar{v}_\pm(\alpha_+,v_w)$. Then the fluid velocity just in front of the bubble wall in the background plasma frame $v_+=\mu(v_w, \bar{v}_+)$ can be known from $\bar{v}_+(\alpha_+,v_w)$. Therefore, the fluid velocity profile $v(\xi)$ can be solved from Eq.~\eqref{eq:dv} with the inital condition chosen at $(v_w, v_+)$. However, the deflagration wave profile should be chopped off at the shockwave front, where $v(\xi)$ suddenly drops to zero at some $\xi=v_{sh}$. 

To determine $v_{sh}$, we first introduce the over-tilde symbol for the shockfront frame, and subscripts `L/R' for the self-similar positions just in the back/front of the shockwave front, respectively, then the junction conditions at the shockfront reads
\begin{align}
w_L\tilde{v}_L\tilde{\gamma}_L^2&=w_R\tilde{v}_R\tilde{\gamma}_R^2,\label{eq:junctionshock1}\\
w_L\tilde{v}_L^2\tilde{\gamma}_L^2+p_L&=w_R\tilde{v}_R^2\tilde{\gamma}_R^2+p_R,\label{eq:junctionshock2}
\end{align}
where $w_L=e_L+p_L=e_R+p_R=w_R$ since the shockwave itself always lives in the false vacuum. This would result in $\tilde{v}_L\tilde{v}_R=1/3$. Since the fluid velocity in the front the shockwave front is at rest, $v_R=\mu(v_{sh},\tilde{v}_R)=0$, leading to the shockfront velocity $v_{sh}=\tilde{v}_R$, which is given previously by $\tilde{v}_R=1/(3\tilde{v}_L)$. Therefore, $v_{sh}$ can be solved from $1/3=v_{sh}\tilde{v}_L=v_{sh}\mu(v_{sh},v_L)$, where $v_L=v(\xi=v_{sh})$ is given by the deflagration profile $v(\xi)$ solved previously from the initial condition at $(v_w, v_+)$.

The enthalpy profile can be similarly obtained from Eq.~\eqref{eq:wxi} by specifying $\xi_0=v_{sh}$ and $w(\xi_0)=w_L$, where $w_L$ is the enthalpy just behind the shockwave front given by
\begin{align}\label{eq:DeflagwL}
w_L=w_R\frac{\tilde{v}_R\tilde{\gamma}_R^2}{\tilde{v}_L\tilde{\gamma}_L^2}=w_N\frac{v_{sh}}{1-v_{sh}^2}\frac{1-\mu(v_{sh},v(v_{sh}))^2}{\mu(v_{sh},v(v_{sh}))}
\end{align}
from the first junction condition~\eqref{eq:junctionshock1} with $w_R=w_N$, $\tilde{v}_R=v_{sh}$, and $\tilde{v}_L=\mu(v_{sh},v(v_{sh}))$. Evolving above enthalpy profile $w(\xi)$ backward from $v_{sh}$ to $v_w$, there is a enthalpy dropping around $v_w$, which can be given by
\begin{align}
w_-=w_+\frac{\bar{v}_+\bar{\gamma}_+^2}{\bar{v}_-\bar{\gamma}_-^2}=w(v_w)\frac{\bar{v}_+}{1-\bar{v}_+^2}\frac{1-v_w^2}{v_w}
\end{align}
from the first junction condition~\eqref{eq:junction1} with $\bar{v}_-=v_w$, $w_+=w(\xi=v_w)$, and $\bar{v}_+(\alpha_+,v_w)$ is already known as a function of $\alpha_+$ and $v_w$. The temperature profile can also be obtained from Eq.~\eqref{eq:Txi} by specifying $\xi_0=v_{sh}$ and $T(\xi_0)=T_L$, where $T_L$ is the temperature just behind the shockfront given by
\begin{align}
T_L=\left(w_L/w_N\right)^{1/4}T_R
\end{align}
from the enthalpy~\eqref{eq:DeflagwL}. Evolving above temperature profile $T(\xi)$ backward from $v_{sh}$ to $v_w$, there is also a temperature dropping around $v_w$, which can be given by
\begin{align}
T_-=\left(\frac{a_+/a_-}{r(\alpha_+,v_w)}\right)^{1/4}T(v_w)
\end{align}
from $r=(a_-T_-^4)/(a_+T_+^4)$ with $T_+=T(\xi=v_w)$.

\subsubsection{Hybrid}

The hybrid mode is a special deflagration mode ($\bar{v}_+<\bar{v}_-$) of Jouguet type ($\bar{v}_-=c_s$) with the bubble wall velocity lying between the sound speed $c_s$ and Jouguet velocity $v_J$ (otherwise it becomes either weak deflagration or weak detonation as shown previously), which contains both compressive shockwave and rarefraction wave in the front and back of the bubble wall, respectively. The Jouguet condition $c_s=\bar{v}_-(\alpha_+,r)$ leads to $r(\alpha_+)$, which in turns determines $\bar{v}_\pm(\alpha_+,r(\alpha_+))\equiv\bar{v}_\pm(\alpha_+)$ and $v_\pm=\mu(v_w,\bar{v}_\pm(\alpha_+))$. Choosing the initial conditions at $(v_w, v_+)$ and $(v_w, v_-)$ just in the front and back of the bubble wall, we can solve from the Eq.~\eqref{eq:dv} the fluid velocity profile $v(\xi)$ forward and backward, respectively. The backward profile vanishes at $\xi=c_s$ just as the detonation mode, and the forward profile will be chopped off at the shockfront $(v_{sh}, v(v_{sh}))$ just as the deflagration mode.

The enthalpy profile can be similarly obtained backward and forward from Eq.~\eqref{eq:wxi} by specifying $(\xi_0,w(\xi_0))=(v_w,w_-)$ and $(\xi_0, w(\xi_0))=(v_{sh},w_L)$, respectively, where the enthalpy $w_L$ just behind the shockfront reads
\begin{align}\label{eq:HybridwL}
w_L=w_R\frac{\tilde{v}_R\tilde{\gamma}_R^2}{\tilde{v}_L\tilde{\gamma}_L^2}=w_N\frac{v_{sh}}{1-v_{sh}^2}\frac{1-\mu(v_{sh}, v(v_{sh}))^2}{\mu(v_{sh}, v(v_{sh}))}
\end{align}
from the first junction condition~\eqref{eq:junctionshock1} with $w_R=w_N$, $\tilde{v}_R=v_{sh}$, and $\tilde{v}_L=\mu(v_{sh}, v(v_{sh}))$, and the enthalpy $w_-$ just behind the bubble wall reads
\begin{align}
w_-=w_+\frac{\bar{v}_+\bar{\gamma}_+^2}{\bar{v}_-\bar{\gamma}_-^2}=w(v_w)\frac{\bar{v}_+}{1-\bar{v}_+^2}\frac{1-c_s^2}{c_s}
\end{align}
from the first junction condition~\eqref{eq:junction1} with $w_+=w(\xi=v_w)$, $\bar{v}_+=\bar{v}_+(\alpha_+)$, and $\bar{v}_-=c_s$.
The temperature profile can also be obtained backward and forward from Eq.~\eqref{eq:Txi} by specifying $(\xi_0, T(\xi_0))=(v_w, T_-)$ and $(\xi_0, T(\xi_0))=(v_{sh}, T_L)$, where the temperature $T_L$ just behind the shockfront reads
\begin{align}
T_L=(w_L/w_N)^{1/4}T_R
\end{align}
from the enthalpy~\eqref{eq:HybridwL}, and the temperature $T_-$ just behind the bubble wall reads
\begin{align}
T_-=\left(\frac{a_+/a_-}{r(\alpha_+)}\right)^{1/4}T(v_w)
\end{align}
from  $r=(a_-T_-^4)/(a_+T_+^4)$ with  $T_+=T(\xi=v_w)$.

\bibliographystyle{utphys}
\bibliography{ref}

\end{document}